\documentclass[journal,twoside,web]{ieeecolor}
\usepackage{jsen}
\usepackage{cite}
\pdfoutput=1 
\usepackage{amsmath,amssymb,amsfonts}
\usepackage{algorithmic}
\usepackage{graphicx}
\usepackage{textcomp}
\usepackage{wrapfig}
\def\BibTeX{{\rm B\kern-.05em{\sc i\kern-.025em b}\kern-.08em
    T\kern-.1667em\lower.7ex\hbox{E}\kern-.125emX}}
\definecolor{abstractbg}{rgb}{0.89804,0.94510,0.83137}
\usepackage{pgfplots} 
\pgfplotsset{compat=1.18, width=3cm}
\usepackage{tikz}
\usepackage{subcaption}
\usetikzlibrary{positioning} 
\usepackage{tikz-3dplot} 

\usepackage{lipsum}

\colorlet{veccol}{green!50!black}
\colorlet{projcol}{blue!70!black}
\colorlet{myblue}{blue!80!black}
\colorlet{myred}{red!90!black}
\colorlet{mydarkblue}{blue!50!black}
\usetikzlibrary{angles,quotes} 
\usepackage{pdfpages}
\usepackage[english]{babel}

\usepackage{etoolbox}

\usepackage{array}
\usepackage[a4paper, total={7.2in, 9in}]{geometry}

\begin{document}
\title{Toward Non-contact Muscle Activity Estimation using FMCW Radar}
\author{Kukhokuhle Tsengwa, \IEEEmembership{}Stephen Paine, Fred Nicolls, Yumna Albertus and Amir Patel \IEEEmembership{}
\thanks{K. Tsengwa, was with the University of Cape Town, Cape Town, South Africa. He is 
now with the Oxford Robotics Institute, Department of Engineering Science, University of Oxford, Oxford, UK (e-mail: kukhokuhle@oxfordrobotics.institute).}
\thanks{S. Paine is with the Radar and Remote Sensing Group (RRSG), Department of Electrical Engineering, University of Cape Town, Cape Town, South Africa (e-mail: stephen.paine@uct.ac.za).}
\thanks{F. Nicolls is with the Digital Image Processing Laboratory, Department of Electrical Engineering, University of Cape Town, Cape Town, South Africa (e-mail: fred.nicolls@uct.ac.za).}
\thanks{Y. Albertus is with the Health, Physical Activity, Lifestyle and Sport (HPALS) Research Centre, Department of Human Biology, University of Cape Town, Cape Town, South Africa (e-mail: yumna.albertus@uct.ac.za).}
\thanks{A. Patel is with the Department of Electrical Engineering, University of Cape Town, Cape Town, South Africa (e-mail: amir.patel@uct.ac.za).}}

\IEEEtitleabstractindextext{%
\fcolorbox{abstractbg}{abstractbg}{%
\begin{minipage}{\textwidth}%
\begin{wrapfigure}[12]{r}{3in}%
\includegraphics[width=3in]{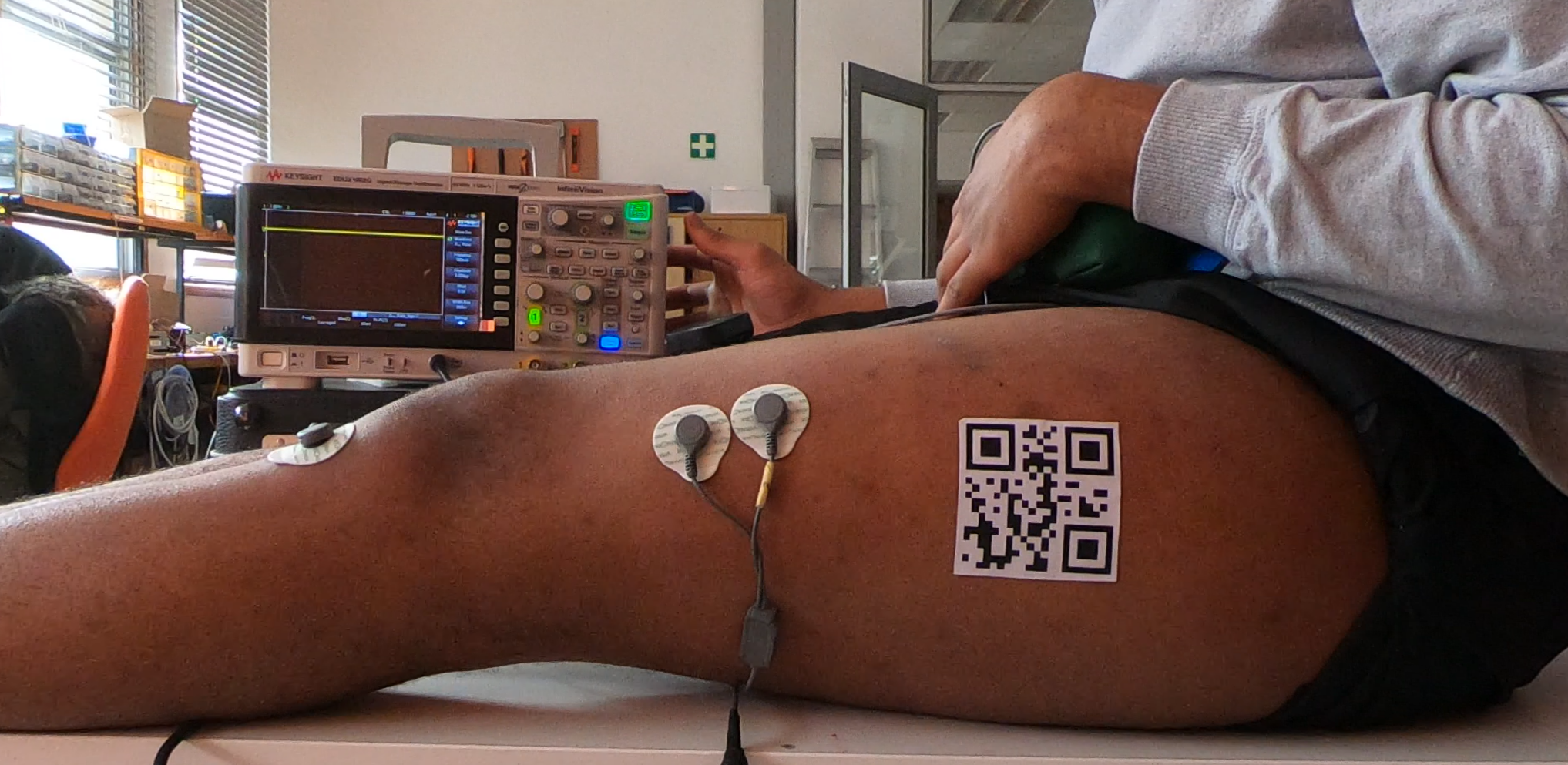}%
\end{wrapfigure}%
\begin{abstract} Surface electromyography (sEMG) is a widely used muscle activity monitoring technique. sEMG measures muscle activity through monopolar and bipolar, multi-electrode electrodes. The surface electrodes are placed on the surface of the skin above the target muscle and the received signal can be used to infer the state of the muscle --- active, inactive or fatigued --- which serves as vital information during neurological and orthopaedic rehabilitation. Additionally, the sEMG signal can also be used for the control of prostheses. sEMG requires contact with the participant’s skin and is thus a potentially uncomfortable method for the measurement of muscle activity. Moreover, the setup procedure has been termed
time-consuming by sEMG experts and is listed as one of the main barriers to the clinical employment
of the technique. Previous studies have shown that architectural changes, particularly muscle deformation, can provide information about the activity of the muscle, providing an alternative to sEMG. In all these studies, the muscle deformation signal is acquired using ultrasound imaging, an approach known as sonomyography (SMG). Despite its advantages, such as improved spatial resolution, SMG is still a contact based approach. In this paper, we propose a non-contact muscle activity monitoring approach that measures the muscle deformation signal using a Frequency Modulated Continuous Wave (FMCW) mmWave radar which we call radiomyography (RMG). In future, this system will enable muscle activation to be measured in an unconstrained and less cumbersome manner for both the person conducting the test and the individual being tested.
\end{abstract}

\end{minipage}}}

\maketitle

\section{Introduction}
\label{sec:introduction}
\IEEEPARstart{M}{uscle} activity monitoring is instrumental to clinicians for neurological and orthopaedic rehabilitation~\cite{Chowdhury2013-yx}~\cite{Woodward2019-cz}. It is also vital as a tool for extracting the signals required for the command of prostheses~\cite{Chowdhury2013-yx}~\cite{8462887}~\cite{Hallock2021-au}~\cite{9707638}. Moreover, monitoring muscle activity required for a particular movement aids in the study of that movement, either to better understand the movement, to inform efforts to replicate the movement in robots~\cite{9025819} or to create anthropomorphic robotic-assisted locomotion for disabled people~\cite{9707638}. \\

Potential difference~\cite{Merletti2016-ez}, sound~\cite{Orizio1993-vg}~\cite{Barry1985}~\cite{DBLP:conf/memea/CasacciaSCTR15}, vibration~\cite{DBLP:conf/memea/CasacciaSCTR15}~\cite{Rohrbaugh2013-mz} and dimensional changes~\cite{8462887}~\cite{Rohrbaugh2013-mz} are the known physical properties whose measurement allows the traditional estimation of muscle activity. Electromyography (EMG) is a muscle activity monitoring approach that measures the potential difference across the cell membranes of muscle cells~\cite{Merletti2016-ez}. Acoustic myography~\cite{Barry1985} and mechanomyography (MMG)~\cite{Woodward2019-cz}~\cite{Rohrbaugh2013-mz} are the names given to the approaches that measure the sound and vibration that occur during muscle activity, respectively. Approaches that measure the dimensional changes of muscles do not yet have an adopted name. Dimensional changes, as measured through ultrasound images, have been referred to as sonomyography (SMG)~\cite{Shi2008-ro}~\cite{Kamatham2022-fc}~\cite{s22072789}. However, this name says more about the sensor used than the physical property measured. \\

\begin{figure}[ht]
    \centering
    \subfloat[\centering]
    {{\includegraphics[width=4.3cm]{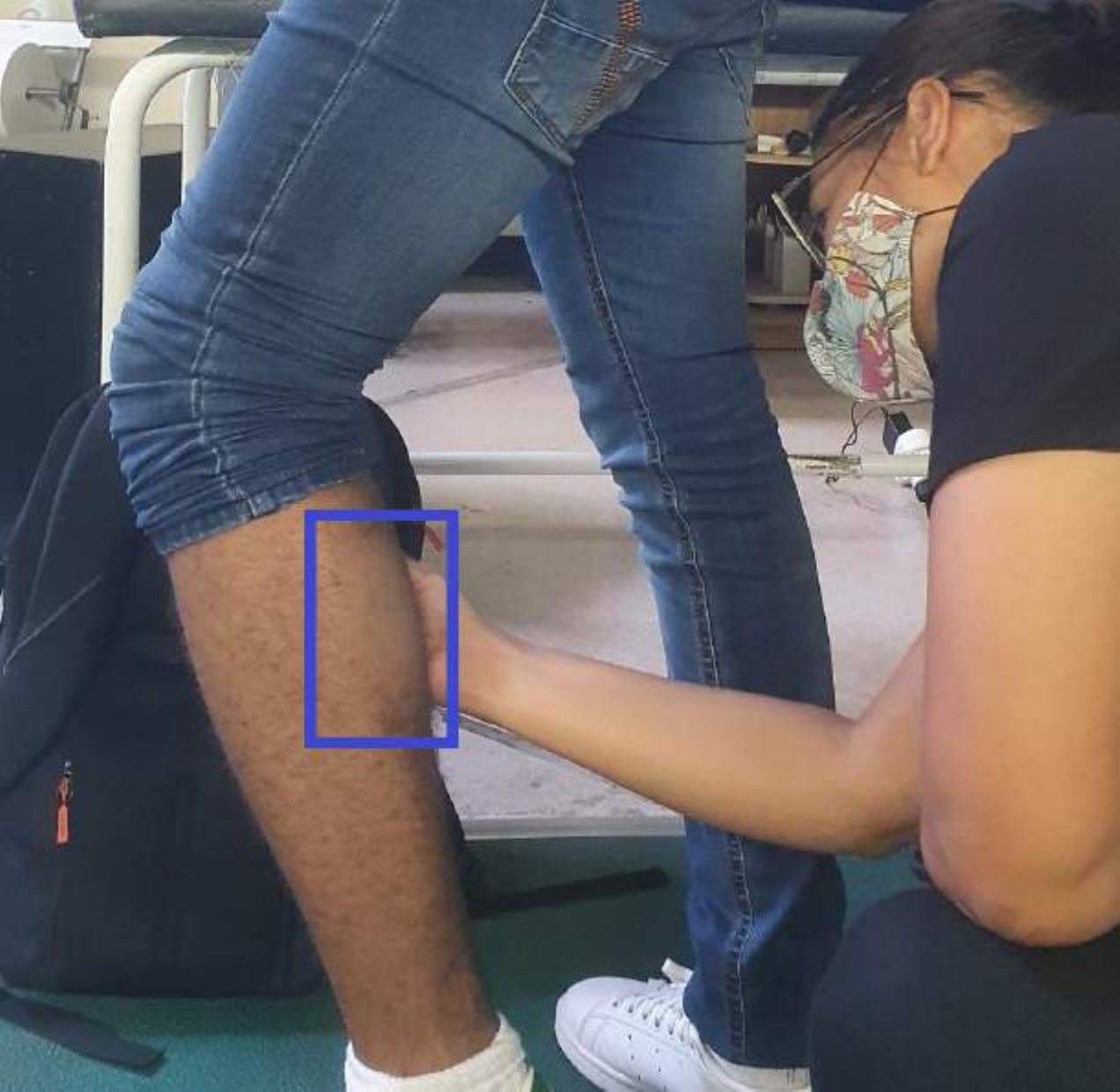}}}
    \subfloat[\centering]
    {{\includegraphics[width=4.3cm]{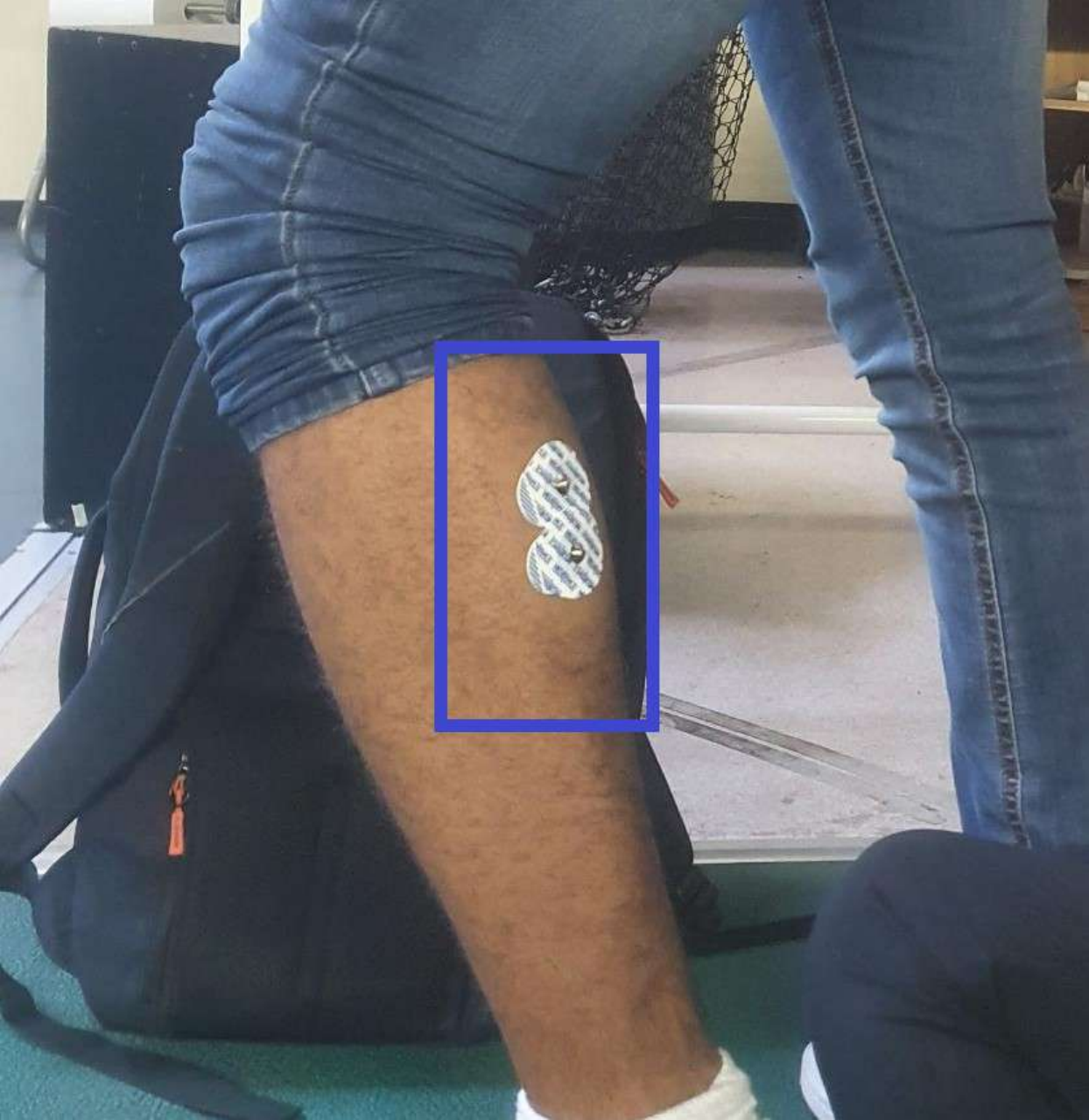}}}
    \caption[The skin over the muscle whose activity is to be monitored with EMG must be prepared.]{ (a) Patch of skin over the gastrocnemius muscle shaved and cleaned with alcohol, and (b) EMG electrodes placed over the lateral gastrocnemius muscle.}
    \label{fig:EMG_Skin_Prep}
\end{figure}

EMG is the current state-of-the-art in muscle activity monitoring. Electrical signals, known as action potentials, from the brain travel through motor neurons and cause the muscle fibers innervated by these neurons to be active~\cite{Merletti2016-ez}. EMG thus monitors muscle activity by measuring these electrical signals through surface electrodes placed on the skin over the muscle of interest (see Fig.~\ref{fig:EMG_Skin_Prep}) or through needle electrodes inserted into the
muscle~\cite{Woodward2019-cz}~\cite{Merletti2016-ez}. EMG is thus a contact based approach for the measurement of muscle activity. \\

Prior to placing the electrodes, the skin was prepared to reduce the impedance and noise at the electrode-skin interface. Skin preparation involves shaving the skin to remove hair, as presented in Fig.~\ref{fig:EMG_Skin_Prep}a. Thereafter, the skin is sometimes rubbed with ethyl alcohol or abrasive conductive paste, washed with water and soap or stripped with adhesive tape to remove oily substances and the most superficial portion of the epidermis. Rubbing with abrasive conductive paste and rinsing is the most effective known treatment for reducing both the noise and impedance at the electrode-skin interface~\cite{Merletti2016-ez}~\cite{seniamWelcomeSENIAM}.\\

As detailed above, EMG requires contact with the participant's skin and is thus an invasive and potentially uncomfortable method for the measurement of muscle activity. Because of tissues between the muscle and electrodes, the EMG signal may be contaminated by muscle activity from other muscles near the muscle of interest. This is called crosstalk~\cite{Merletti2016-ez}. Moreover, the setup procedure has been termed time-consuming by EMG experts and is listed as one of the main barriers to the clinical employment of the technique~\cite{Manca2020-ez}.\\

The low amplitude of the sound produced during muscle activity limits acoustic myography to being a contact based approach. Microphones are often used to measure this signal~\cite{DBLP:conf/memea/CasacciaSCTR15}~\cite{Rohrbaugh2013-mz}. Vibration and dimensional changes, however, are properties that can potentially be measured without contact. Indeed, a Laser Doppler Vibrometer (LDV) has been used by Rohrbaugh et al. to measure the vibrations that accompany muscle activity (the MMG signal), in a technique known as Laser Doppler Myography (LDMi)~\cite{DBLP:conf/memea/CasacciaSCTR15}~\cite{Rohrbaugh2013-mz}~\cite{517ff2062ffd4c67869d303afd50d7f7}. Dimensional changes as measured through SMG would be technically impractical to measure without contact because of the large difference between the acoustic impedance of air (\(0.4\times10^3\)~kg\(\cdot\)m\(^{-2}\cdot\)s\(^{-1}\)~\cite{REGTIEN2018267}) and that of human skin (\(1.99\times10^{6}\)~kg\(\cdot\)m\(^{-2}\cdot\)s\(^{-1}\)~\cite{Haim_Azhari_2010-qo}), which would lead to most of the
acoustic energy being reflected at the air-skin interface.\\

It has been shown in previous studies~\cite{Shi2008-ro}~\cite{9224391}~\cite{Kamatham2022-fc}~\cite{Hallock2021-au} that dimensional or architectural changes in a muscle during its contraction correlates well with the muscle's activity and output force. These informative architectural parameters of muscles include muscle thickness, shape, cross-sectional area, muscle fiber pennation angle and the position of the muscle's surface under the skin. When a muscle contracts, its sarcomeres (the basic contractile units of muscles) necessarily change length~\cite{Merletti2016-ez}~\cite{Hallock2021-au}. Because the volume of muscles remains constant~\cite{McMahon1984-rk}, it
follows that muscle activity leads to shape change or deformation of the muscle~\cite{Hallock2021-au}. It has been
demonstrated that this deformation correlates with muscle activity~\cite{Shi2008-ro}~\cite{9224391}~\cite{Kamatham2022-fc}~\cite{Hallock2021-au}. In most studies, this muscle deformation signal has been measured and the underlying muscle activity monitored by measuring either muscle cross-sectional area, thickness~\cite{Hallock2021-au} or muscle surface position~\cite{Shi2008-ro} across a sequence of ultrasound images of the muscle. This approach is known as sonomyography (SMG). \\

SMG offers improved spatial resolution as compared to EMG, which solves the problem of muscle crosstalk~\cite{Shi2008-ro} and can
measure activity from deep seated muscles~\cite{Kamatham2022-fc}. However, most SMG algorithms make use of computationally expensive tools such as optical flow to recover the SMG signals~\cite{9224391}~\cite{Hallock2021-au}. There is no non-contact muscle activity monitoring approach that measures the characteristic muscle deformation signal seen
in the SMG literature. In this paper, we address this gap.\\

Frequency Modulated Continuous Wave (FMCW) is a specific type of modulation scheme used for
active sensors. It is often implemented using radio waves with wavelengths in the millimetre range, a technique known as mmWave FMCW radar. The relatively short wavelengths, compared to traditional radars, allow mmWave radars to be compact and low-cost~\cite{Chen2019-wh}. Additionally, the high frequencies used (60-64\,GHz) allow for wide bandwidths (up to 4\,GHz) and thus fine range resolution. The FMCW radar signal is transmitted as frequency modulated sweeps known as chirps. This makes FMCW radars particularly suitable to the monitoring of small motions by analysing the change in phases across the consecutive chirps. \\

Owing to the properties above, FMCW radar has been widely and successfully used to measure human vital signs such as heart and breathing rate without contact~\cite{heartRateSensingICRA2020}~\cite{Adib2015-ld}. With FMCW radar, heart and breathing rates have been measured with a median accuracy of \(99\%\)~\cite{Adib2015-ld}. The non-contact measurement of these signals using FMCW radar is made possible by the movement of the chest in response to the beating of the heart and the inflation and deflation of the lungs. The muscle activity monitoring approach presented in this study was
inspired by this prior work. \\

We propose a non-contact muscle activity monitoring approach that measures the muscle deformation signal using an off-the-shelf mmWave FMCW radar. Our muscle activity monitoring approach measures the small motion at the surface of the skin over the muscle of interest. Our approach measures the same physical attribute measured by sonomyography (SMG) and we have therefore dubbed it radiomyography (RMG). From data collected during experiments to evaluate the performance of our system, we observed effects and phenomena that are characteristic of muscles and well known in biomechanics. \\

\section{Frequency Modulated Continuous Wave (FMCW) Radar Theory}

As the name suggests, the frequency of the transmitted waveform in an FMCW radar is modulated. A sawtooth modulation scheme, known as a chirp, is often used~\cite{Chen2019-wh}. This means that the frequency of the transmitted waveform, \(f_T(t)\), varies linearly with time according to:

\begin{equation} \label{transmit_chirp_freq_eqn}
f_T(t) = \frac{B}{T}t + f_0,
\end{equation}

where \(B\) is the bandwidth of the frequency sweep, \(T\) is the time it takes to sweep the bandwidth and \(f_{0}\) is the starting frequency. This frequency variation across time is why the transmitted signal is called a chirp. \\

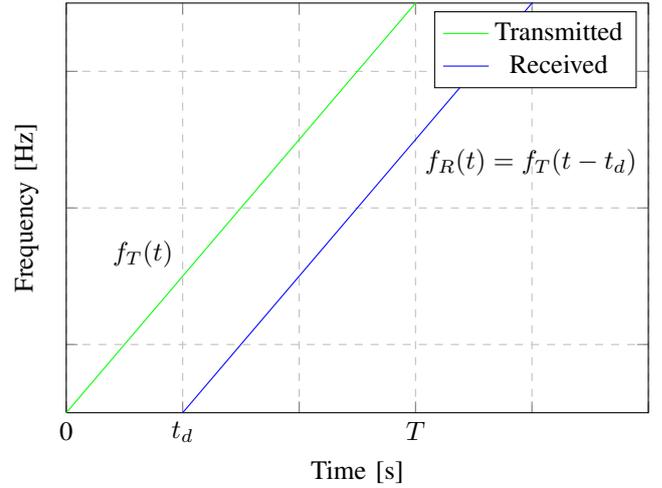
\begin{figure}[ht]
    \centering
    
    \begin{tikzpicture}
    \begin{axis}[title={Frequency of chirp transmitted by an FMCW radar and the frequency of the corresponding received chirp},
    title style={text width=10cm, align=center},
    clip=true, yticklabels={,\(f_o\),,,,,,\(f_o + B\)}, xticklabels={,0,\(t_d\),,\(T\)}, xmin=0, ymax=700, ymin=100, xmax=5, xlabel={Time [s]}, ylabel={Frequency [Hz]}, xmajorgrids=true, grid style=dashed, ymajorgrids=true, grid style=dashed, scale=5.4]
    
    \addplot[color=green]{200*x + 100};
    \filldraw[black] (1,300) circle (0pt) node[anchor=south east]{\(f_T(t)\)};
    \addlegendentry{Transmitted}
    
    \addplot[color=blue]{200*(x-1) + 100};
    \filldraw[black] (3,500) circle (0pt) node[anchor=north west]{\(f_R(t) = f_T(t-t_d)\)};
    \addlegendentry{Received}
    
    \end{axis}
    \end{tikzpicture}
    
    \caption[Frequency of a typical chirp transmitted by an FMCW radar and the frequency of the corresponding reflected (or received) chirp.]{Frequency of a typical chirp transmitted by an FMCW radar and the frequency of the chirp received by the radar time \(t_d\) after the chirp was trasmitted. The transmitted chirp reflected off a target a distance \(R_0\) from the radar.}
    \label{fig:transmit_receive_freqs}

\end{figure}

As seen in~\cite{8123923}, a typical chirp transmitted by an FMCW radar can be represented as
 
 \begin{equation} 
\label{transmitted_chirp_eqn}
S_T(t) = A_Te^{j(2\pi f_0t + \pi K t^2) + \phi_0},  0\leq t\leq T,
\end{equation}

where \(K = \frac{B}{T}\) is the slope of the transmitted frequency, \(\phi_0 \) is the initial phase and \(A_T\) is the amplitude of the transmitted chirp.\\

\begin{figure*}[ht]
    \centering
    \includegraphics[scale=0.6]{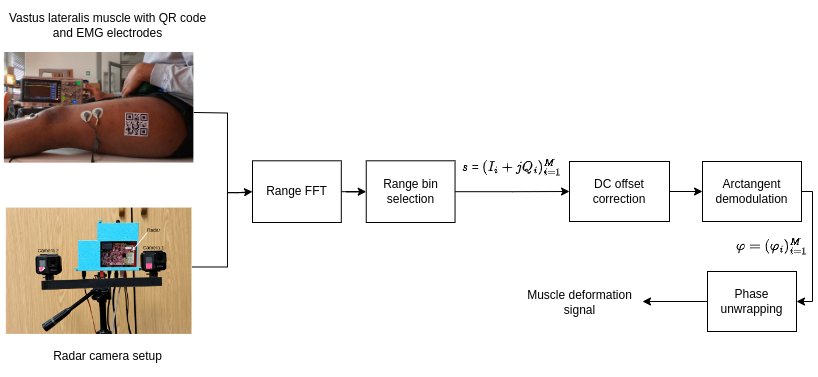}
    \caption{A flow diagram detailing each stage of the advanced signal processing techniques, from receiving the radar and camera signals to outputting the muscle movement.}
    \label{fig:Camera Based Phase Estimation Flowchart}
\end{figure*}

This chirp is transmitted and reflects off a target at a nominal distance of $R_0$ from the radar. Fig.~\ref{fig:transmit_receive_freqs} shows the linearly varying frequency of the transmitted chirp, \(f_T(t)\), as well as that of the reflected (or received) chirp, \(f_R(t)\). Let \(x(t)\) be the component of the target's displacement along the radar's boresight. Then the received chirp, \(S_R(t)\), after time \(t_d\) becomes

 \begin{equation} 
\label{received_chirp_eqn}
S_R(t) = S_T(t-t_d) = A_Re^{j(2\pi f_0(t-t_d) + \pi K (t - t_d)^2) + \phi_0},
\end{equation}

where \(A_R\) is the amplitude of the received chirp and the time delay, \(t_d\), is given by

\begin{equation} \label{round_trip_delay_time_eqn2}
t_d = \frac{2(R_0 + x(t))}{c},
\end{equation}

where \(c\) is the speed of electromagnetic radiation.\\

The amplitude of the received chirp, \(A_R\), depends on multiple factors such as the peak transmit power of the transmitter, transmit antenna gain, receiver antenna gain, radar cross section (RCS) of the target, the distance between the target and the radar, etc.~\cite{Chen2019-wh}.\\

In the receiver stage, the signal \(S_R(t)\) is amplified by a low noise amplifier (LNA), before being mixed with the original transmit chirp to create the baseband signal known as the beat frequency, before passing through low-pass anti-aliasing filter~\cite{Chen2019-wh}~\cite{8123923}. A second intermediate frequency (IF) amplification stage is then used before the signal gets digitised. The beat frequency that is digitised, \(S_B(t)\), in~\eqref{received_chirp_eqn2}~\cite{8123923}, can be represented as an in-phase component (I) and a quadrature component (Q): 

 \begin{equation} 
\label{received_chirp_eqn2}
S_B(t) = S_T(t) \cdot S_R(t)^\ast = A_Re^{j(2\pi Kt_dt + 2\pi f_0  t_d) },
\end{equation}

where \((\cdot)^\ast\) is the complex conjugate operator.\\

The term associated with $t_d^2$ has been ignored since $t_d^2 \ll t_dt$~\cite{8123923}. Additionally, it has been assumed that during the chirp duration, \(T\), the displacement of the target is negligible and thus the target has no velocity during this time~\cite{8123923}. If this assumption is not made (or is unreasonable), then the frequencies in both \(S_R(t)\) and \(S_B(t)\) should have a Doppler frequency shift term added to them~\cite{Chen2019-wh}~\cite{8123923}.\\

Now let us assume that the radar transmits and receives a total of \(M\) chirps, each with index $i \in \{1, 2, 3, ..., M\}$. Also suppose that each received chirp has N samples. Then any received chirp can be generally described by

\begin{equation} \label{general_received_chirp_eqn}
\begin{split}
S_B(iT + t) & = A_Re^{j(\frac{4\pi K R_0}{c}t + \frac{4\pi f_0  R_0}{c} + \frac{4 \pi f_c}{c}x(iT)) }  \\ 
 & = A_Re^{j(2\pi f_b t + \frac{4\pi f_0  R_0}{c} + \frac{4 \pi f_c}{c}x(iT)) } \\
 & = A_Re^{j(2\pi f_b t + \frac{4\pi  R_0}{\lambda_0} + \frac{4 \pi x(iT)}{\lambda_c}) } \\
 & =  A_Re^{j(2\pi f_b t + \varphi_i) },
\end{split}
\end{equation}

where \(f_c = f_0 + \frac{B}{2}\) is the center frequency, \(\lambda_c \) is the center wavelength, \(f_b = \frac{2BR_0}{cT}\) is the beat frequency and \(\varphi_i = \frac{4\pi  R_0}{\lambda_0} + \frac{4 \pi x(iT)}{\lambda_c}\).\\

It is thus clear that the phase of each received chirp, $\varphi_i$, encodes within it a sample of the small motion at time $iT$, $x(iT)$. Just as importantly, we see that the small motion is sampled at frequency $\frac{1}{T}$. Because of the Nyquist-Shannon sampling criterion, this frequency (or equivalently, period) places an upper limit on how fast the target can move while still being unambiguous with this approach. \\

A challenge that faces any approach that monitors a target's small motion is the presence of larger
motions that corrupt the signal of interest. Consider that the target exhibits not only the motion of interest, \(x(t)\), but also some other motion, \(\rho(t)\), then the recovered phase would be \(\varphi_i = \frac{4\pi  R_0}{\lambda_0} + \frac{4 \pi x(iT)}{\lambda_c} + \frac{4 \pi \rho(iT)}{\lambda_c}\). This unwanted motion is called random body motion/movement (RBM)~\cite{6662489}. If RBM is of sufficient magnitude, then the signal of interest can be completely masked. 

\section{Muscle Activity Estimation Pipeline}

\begin{figure*}[ht]
    \centering
    \includegraphics[scale=0.45]{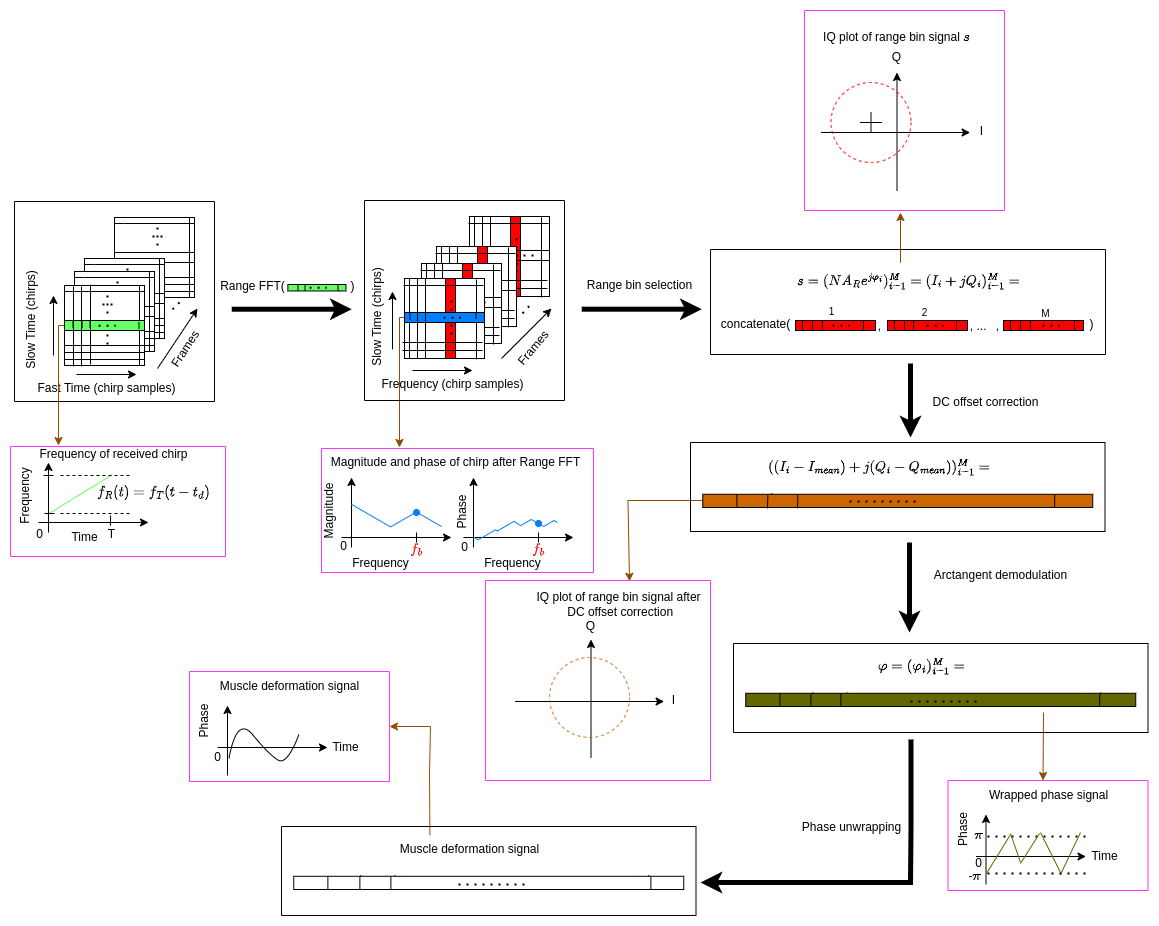}
    \caption{A detailed pictorial description of the operations within each of the stages of our muscle activity estimation pipeline. }
    \label{fig:detailedprocessingpipeline}
\end{figure*}

Fig.~\ref{fig:Camera Based Phase Estimation Flowchart} above depicts the different stages in our signal processing pipeline. For a pictorial description of each stage refer to Fig.~\ref{fig:detailedprocessingpipeline}. Each stage is described below. Firstly, a Discrete Fourier Transform (DFT) was applied on each of the \(M\) received chirps at baseband. The frequency axis of this DFT can be interpreted as range and the magnitude of the Fourier
coefficients are a measure of how much energy is received from that range. For this reason, this operation is called the Range Fast Fourier Transform (Range FFT). \\


Mathematically, this is 

\begin{equation} 
\label{DFT_eqn1}
X_k = \sum_{n=0}^{N-1} x_n e^{-j2 \pi \frac{k}{N} n},
\end{equation}

where \(k\) is the frequency index, \(x_n = S_B(iT + n\cdot T_s), n\in \{0, 1, 2, ..., N-1\} \) is the discretised received chirp, \(N\) is the number of samples per received chirp and \(T_s = \frac{1}{f_s}\) is the period at which the received chirp has been sampled.\\

Dividing and multiplying by \(T_s\) in~\eqref{DFT_eqn1} gives

\begin{equation} 
\label{DFT_eqn2}
X_k = \sum_{n=0}^{N-1} x_n e^{-j2 \pi \frac{k}{NT_s} n T_s}.
\end{equation}

If we let \( \frac{k}{N T_s} = \frac{k}{N}f_s = f_k  \) and substitute into~\eqref{DFT_eqn2} the correct expression for \(x_n\), we get

\begin{equation} 
\label{DFT_eqn3}
\begin{split}
X_{f_{k}}   & = \sum_{n=0}^{N-1} A_R e^{j2\pi f_b n T_s} e^{-j2 \pi f_k  n T_s} e^{j \varphi_i}\\
 & = A_Re^{j \varphi_i} \sum_{n=0}^{N-1} e^{j2\pi (f_b-f_k ) n T_s}. \\
\end{split}
\end{equation}

If \(f_k =f_b\), then 

\begin{equation} 
\label{DFT_eqn4}
\begin{split}
X_{f_{k}}  = NA_Re^{j \varphi_i} \\
\end{split}
\end{equation}

because \(\sum_{n=0}^{N-1} e^{j2\pi (f_b-f_k) n T_s} = N\). Therefore, at the beat frequency, the Range FFT coefficient has magnitude \(|X_{f_{k}}| = NA_R\).\\

However, if $f_k \neq f_b$, then we can let $z = e^{j2\pi (f_b-f_k) T_s}, z \in \mathbf{C}$. Then by noticing that our sum is a geometric sum, we have

\begin{equation} 
\label{DFT_eqn5}
\begin{split}
\sum_{n=0}^{N-1} e^{j2\pi (f_b-f_k) n T_s} = \sum_{n=0}^{N-1} z^{n} = 1 + z + z^2 + z^3 + ... + z^{N-1}.\\
\end{split}
\end{equation}

Letting $|\sum_{n=0}^{N-1} e^{j2\pi (f_b-f_k) n T_s} | = N^{\ast}$ and applying the triangle inequality on~\eqref{DFT_eqn5} gives

\begin{equation} 
\label{DFT_eqn6}
\begin{split}
\bigg|\sum_{n=0}^{N-1} e^{j2\pi (f_b-f_k) n T_s} \bigg| \leq |1| + |z| + |z^2| + ... + |z^{N-1}| = N.\\
\end{split}
\end{equation}

Therefore, at any frequency that is not the beat frequency, the Range FFT coefficient has magnitude $|X_{f_{k}}| = N^{\ast}A_R$. But~\eqref{DFT_eqn6} states that $N^{\ast} \leq N$. Therefore, the magnitude of the Range FFT coefficient at the beat frequency is maximal i.e. there will not be a Range FFT coefficient with magnitude greater than this one at any other frequency. This is what allows us to determine where the target is i.e. $R_0$, accurate to within the range resolution. Often, in practice, $|X_{f_{k}}| = NA_R$ is not only maximal but is the maximum, which makes finding $R_0$ easier.\\

Finding the beat frequency for the target of interest, or equivalently, its nominal distance from the radar, \(R_0\), is what we refer to as the range bin selection stage of our pipeline. Once the beat frequency is known, we extract the Range FFT coefficient at the beat frequency for all received chirps. We call the sequence of all these coefficients the range bin signal, \(s\). From~\eqref{DFT_eqn4}, the range bin signal is 

\begin{equation} 
\label{range_bin_signal_eqtn}
\begin{split}
s = (NA_Re^{j \varphi_i})_{i=1}^M = (I_i + jQ_i)_{i=1}^M,
\end{split}
\end{equation}

where \(I_i\) and \(Q_i\) are the real and imaginary parts of the \(ith\) Range FFT coefficient, respectively.\\

Recall that at the receiver output, the complex-valued baseband signal, \(S_B(t)\), in~\eqref{received_chirp_eqn2} comprises a real and an imaginary part or an in-phase (I) and a quadrature (Q) component, respectively. Ideally, the I and Q components are sinusoids, with a phase offset of \(90^{\circ}\) between each other and with equal amplitudes~\cite{Chen2019-wh}. In reality, due to imperfections in the receiver's electronics, amplitude and phase imbalances exist between the two channels~\cite{Chen2019-wh}~\cite{imbalanceCorr1}~\cite{imbalanceCorr2}. Additionally, the range bin signal often contains DC offsets due to coupling effects and environmental interferences~\cite{9716146}. Both of these negatively affect the ability to recover the phase signal and thus must be corrected for. \\

Through experimenting with different algorithms, we found that the DC offset correction algorithms~\cite{Chen2019-wh}~\cite{Alizadeh2019RemoteMO} were very effective at improving phase signal retrieval whereas the IQ imbalance algorithms~\cite{Chen2019-wh}~\cite{imbalanceCorr2} had no positive effect. The DC offset correction algorithm we specifically implement here subtracts the mean of the range bin signal from the range bin signal. This is represented by the DC offset correction stage in the pipeline.\\

At the arctangent demodulation stage, the phase signal is obtained from the range bin signal by taking the arctangent of the imaginary part, \(Q_i\), over the real part, \(I_i\), for each sample in the range bin signal~\cite{Chen2019-wh}. At the output of this stage is the phase signal

\begin{equation} 
\label{DFT_eqn7}
\begin{split}
\varphi = (\varphi_i)_{i=1}^M.
\end{split}
\end{equation}

Because the phase samples, \(\varphi_i\), in the range bin signal are in the range \([-\pi, \pi]\) (we say the phase is wrapped), after arctangent demodulation it is often necessary to unwrap the phase. This is necessary because if say\(\ x(iT) = \frac{\lambda_c}{4}\), then \(\varphi_i = \frac{4\pi  R_0}{\lambda_0} +  \pi > \pi\) for \(R_0 \neq 0 \) (see~\eqref{general_received_chirp_eqn}). 

Phase unwrapping works by making the assumption that the target does not exceed the maximum permissible velocity, \(v_{max}\), as determined by parameters set on the radar~\cite{Alizadeh2019RemoteMO}. From~\eqref{general_received_chirp_eqn}, recall that $\varphi_i = \frac{4\pi  R_0}{\lambda_0} + \frac{4 \pi x(iT)}{\lambda_c}$. Notice that if, \(x((i+1)T) = x(iT) + \frac{\lambda_c}{2} \), then $\varphi_{i+1} = \frac{4\pi  R_0}{\lambda_0} + \frac{4 \pi x(iT)}{\lambda_c}+ 2 \pi =  \varphi_i $. Even though the small motion has changed by $\frac{\lambda_c}{2}$ over a time duration of \(2T\), the phases do not capture this motion. Therefore, we conclude that over a time equal to two chirp times, \(2T\), the small motion should change less than \(\frac{\lambda_c}{2} \). This argument actually gives us an expression for the maximum velocity, \(v_{max}\), the target can have before it becomes impossible to encode its small motion in the phase. We have

\begin{equation} 
\label{maximum_velocity_eqtn}
\begin{split}
v_{max} = \frac{ \frac{\lambda_c}{2}  }{2T} = \frac{\lambda_c}{4T} = \frac{  \frac{\lambda_c}{4}}{T}.
\end{split}
\end{equation}

One interpretation of~\eqref{maximum_velocity_eqtn} is that the largest displacement the target can have in a time equal to the chirp time, \(T\), is \(\frac{\lambda_c}{4}\). So, given

\begin{equation}
\label{phase_sample_i_eqtn1}
\begin{split}
\varphi_i = \frac{4\pi  R_0}{\lambda_0} + \frac{4 \pi x(iT)}{\lambda_c} 
\end{split}
\end{equation}

and 

\begin{equation}
\label{phase_sample_i_eqtn2}
\begin{split}
\varphi_{i+1} = \frac{4\pi  R_0}{\lambda_0} + \frac{4 \pi x((i+1)T)}{\lambda_c},   
\end{split}
\end{equation}

we have 

\begin{equation}
\label{phase_sample_i_eqtn3}
\begin{split}
\Delta \varphi = \varphi_{i+1} - \varphi_i = \frac{4 \pi \Delta x}{\lambda_c} 
\end{split},
\end{equation}

where

\begin{equation}
\label{phase_sample_i_eqtn4}
\begin{split}
\Delta x = x((i+1)T) - x(iT)
\end{split}
\end{equation}

Therefore, if the largest value that \(\Delta x\) can attain is \(\frac{\lambda_c}{4}\), then the largest value that \(\Delta \varphi\) can be is \(\pi\). Because the target can move towards and away from the radar, we also consider that the displacement \(\Delta x\) cannot be less than \(-\frac{\lambda_c}{4}\). This then suggests that the smallest \(\Delta \varphi\) can be is \(-\pi\). We have

\begin{equation} 
\label{phase_diff_inequality}
\begin{split}
-\pi \leq \varphi_{i+1} - \varphi_i \leq \pi.
\end{split}
\end{equation}

This result leads to the phase unwrapping algorithm proposed in~\cite{Alizadeh2019RemoteMO} and implemented in the last stage of our pipeline. The algorithm works as follows: iterate over all the wrapped phases. If the difference between two consecutive phases satisfies~\eqref{phase_diff_inequality}, do nothing. If instead $\varphi_{i+1} - \varphi_i > \pi$, then subtract \(2\pi\) from \(\varphi_{i+1}\). Finally, if $\varphi_{i+1} - \varphi_i < -\pi$, then add \(2\pi\) to \(\varphi_{i+1}\). \\

The output of our signal processing pipeline is therefore the unwrapped phase signal, \(\varphi\), which encodes the small motion exhibited by the target at the nominal distance \(R_0\). In our experimental setup, the radar faced the surface of the skin above the muscle of interest. Assuming that the only motion exhibited by the skin was due to the movement of the muscle underneath it, the phase signal encodes the muscle deformation signal which we use as a measure of muscle activity. 


\section{Experiments}
\label{sec:experiments}

To evaluate the performance of our radar based muscle activity monitoring system, we recruited three participants from which the required data was collected. All participants were aged between 22 and 25 years. During each experiment, two sets of time-synchronised data were collected, one from the radar and another from the EMG sensor. The radar and EMG sensor are triggered using an external hardware trigger.\\

\begin{figure}[h]
    \centering
    \includegraphics[scale=0.3]{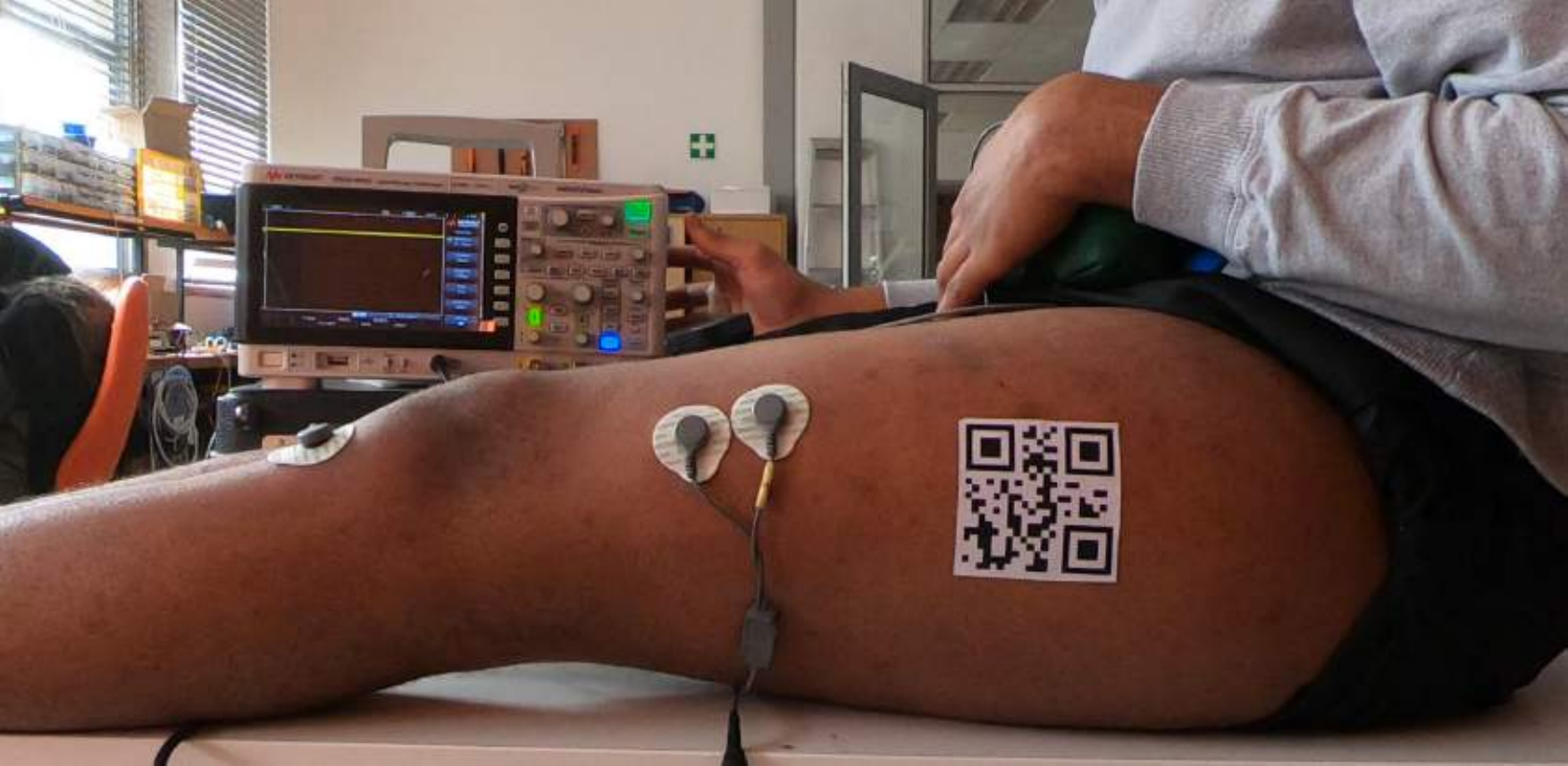}
    \caption[Each participant sits on top of a table that was placed \(70\)~cm in front of the tripod while wearing shorts to expose the QR code to the cameras.]{Each participant sat on top of a table that was placed \(70\)~cm in front of the tripod while wearing shorts to expose the QR code to the cameras.}
    \label{fig:my thigh with emg electrodes and qr code}
\end{figure}

Four experiments, each roughly \(1\)~minute (\(57.1\)~s) long, were conducted on each participant. During an experiment, a participant was asked to isometrically contract and relax their vastus lateralis muscle (the strongest and largest quadriceps muscle that runs along the entire length of the lateral side of the thigh) randomly. A pair of EMG electrodes were placed over the vastus lateralis and one EMG ground electrode was placed over the patellar tendon (see Fig.~\ref{fig:my thigh with emg electrodes and qr code}).

The muscle activity monitoring experimental setup consists of a Texas Instruments~(TI) IWR6843ISK mmWave FMCW radar~\cite{tiIWR6843ISKEvaluation} and the DCA1000EVM data capture board~\cite{tiDCA1000EVMEvaluation}. Both devices were enclosed within a 3D-printed enclosure which was mounted on a 35\,cm long rectangular 3D-printed rod. These two devices were connected to the MMWAVEICBOOST~\cite{tiMMWAVEICBOOSTEvaluation}, which receives the synchronising signal. On either side of the radar was a GoPro HERO~7 Black camera (see Fig.~\ref{fig:My Camera Radar Rig Setup} below) with frame rate and resolution set to \(120\)~fps and \(1920 \times 1080\)~pixels, respectively. We refer to these as camera~1 and 2 and the data from these cameras was used  to corroborate the vastus lateralis’ range estimation from the radar. Both cameras were also mounted on the same rod, and the rod was fixed to an adjustable-height tripod. It is important to note that the cameras are completely unnecessary for the operation of the proposed non-contact muscle activity monitoring system.

\begin{figure}[h]
    \centering
    \includegraphics[scale=0.3]{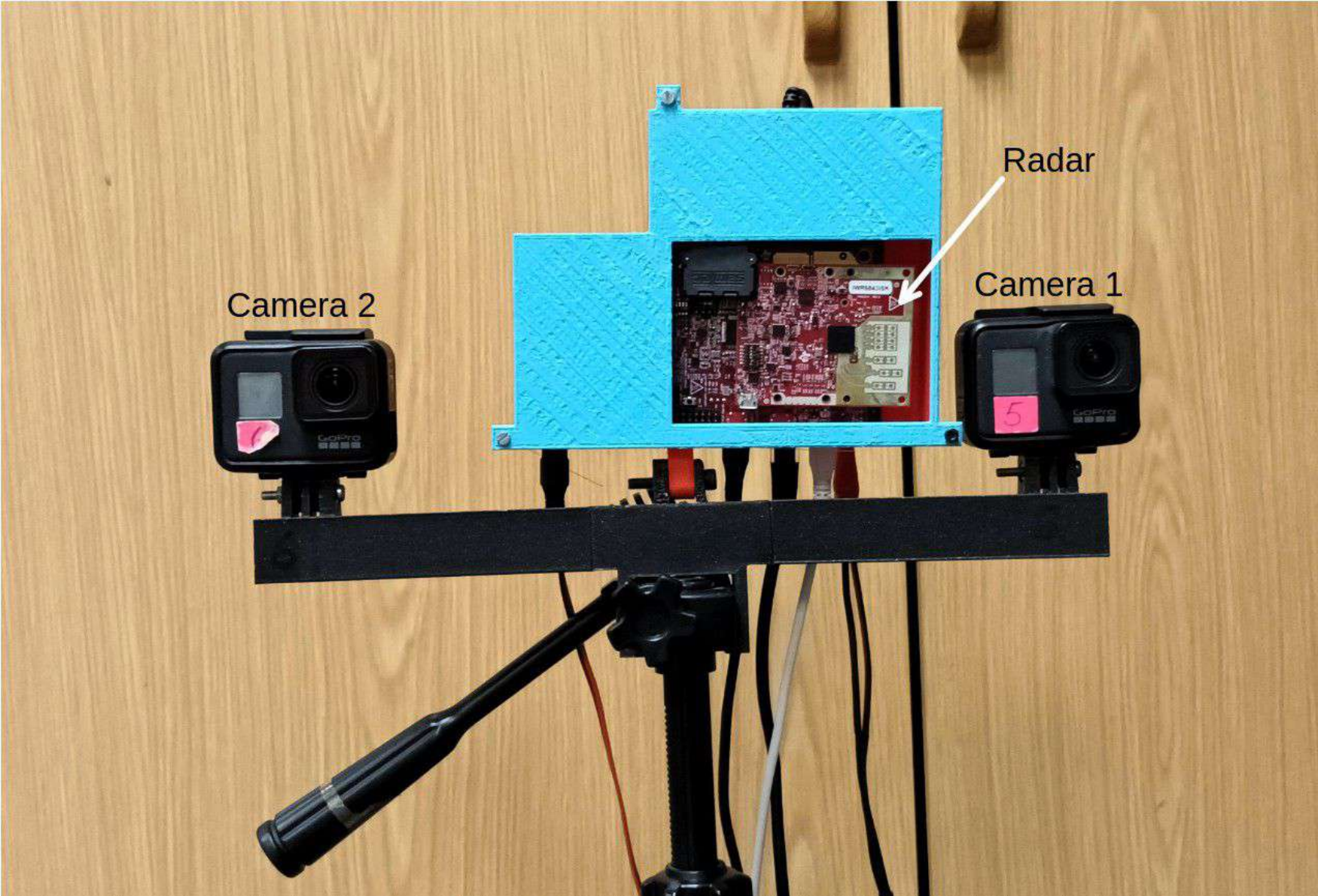}
    \caption[Our muscle activity monitoring experimental setup consists of the Texas Instruments~(TI) IWR6843ISK mmWave FMCW radar and the DCA1000EVM data capture board.]{Our muscle activity monitoring experimental setup consists of the Texas Instruments~(TI) IWR6843ISK mmWave FMCW radar and the DCA1000EVM data capture board. Both devices were enclosed within a 3D-printed enclosure which was then mounted on a 35\,cm long rectangular 3D-printed rod. On either side of the radar was a GoPro HERO~7 Black camera.}
    \label{fig:My Camera Radar Rig Setup}
\end{figure}

 A QR code was also placed on the skin over the vastus lateralis muscle using paper glue. The skin was shaved with a  razor blade for each participant, enabling the QR code to be placed onto the skin surface. The QR code was used together with a calibrated stereo camera pair to estimate how far the vastus lateralis muscle was from the radar using triangulation. This range estimation was used to corroborate the vastus lateralis' range estimation from the radar. Recall that range estimation is equivalent to range bin selection, which was one of the stages in our muscle activity monitoring pipeline. It is to be noted that the radar's range estimation is sufficient and that the QR code is not required at any stage of our pipeline. \\

Each participant sat on top of a table that was placed 70\,cm in front of the tripod while wearing shorts to expose the QR code to the cameras, as presented in Fig.~\ref{fig:my thigh with emg electrodes and qr code}. It should be noted however that the radar can penetrate through clothing. The tripod's height was adjusted such that the radar's boresight was pointing directly at the participant's vastus lateralis muscle.\\

Ethical approval for this study was obtained from the University of Cape Town Faculty of Health Sciences Human Research Ethics Committee (HREC REF: 379/2022).

\section{Results and Discussion}
\label{sec:results_and_discussion}

The results presented here are from a total of 12 experiments across the three
participants. Four experiments were conducted for each participant. Each experiment was about 1~minute (57.1\,s) long. Fig.~\ref{fig:full emg, radar phase and camera phase encoding muscle activity} below illustrates the EMG and radar phase signals recorded from one such experiment, specifically from participant~1. We observe that each time the participant contracts the vastus lateralis, the radar phase signal increases. This implies that contraction deforms the vastus lateralis muscle away from the radar. Conversely, when the muscle was relaxed, the radar phase decreases. This suggests that relaxation deforms the muscle towards the radar. \\

The most noticeable feature of Fig.~\ref{fig:full emg, radar phase and camera phase encoding muscle activity} is the high correlation between the EMG signal and the deformation signal (as measured by the phase signal). Several other studies~\cite{Shi2008-ro}~\cite{9224391}~\cite{Hallock2021-au} have demonstrated that muscle deformation (as measured through ultrasound images) correlates with muscle activity (as measured through EMG) and/or muscle force.\\

\begin{figure}[h]
    \begin{subfigure}[b]{\textwidth}
    \resizebox{0.52\textwidth}{!}{
    \begin{tikzpicture}
    \begin{axis}[title = {EMG signal from left vastus lateralis muscle},
    title style={text width=10cm, align=center},
    grid, 
    grid style={gray!30},
    xlabel = {Time [s]},
    ylabel = {EMG [V]},
    xmin=0,
    xmax=57.1423,
    xtick={0, 5, 10, 15, 20, 25, 30, 35, 40, 45, 50, 55},
    scale=5]
    \addplot[mark size=0.2, grid, color=red]table[y index=1]{participant1Exp1MAEMGDataDownSampled.txt};
    \end{axis}
    \end{tikzpicture}
    }
    \caption{EMG signal from left vastus lateralis muscle}
    \label{fig:full emg envelope signal vs time}
    \end{subfigure} 
    \begin{subfigure}[b]{\textwidth}
    \resizebox{0.52\textwidth}{!}{
    \begin{tikzpicture}
    \begin{axis}[title = {Radar phase signal},
    title style={text width=10cm, align=center},
    grid, 
    grid style={gray!30},
    xlabel = {Time [s]},
    ylabel = {Phase [rad]},
    xmin=0,
    xmax=57.1423,
    xtick={0, 5, 10, 15, 20, 25, 30, 35, 40, 45, 50, 55},
    scale=5]
    \addplot[mark size=0.2, grid, color=blue]table[y index=1]{participant1Exp1MARadarCameraData.txt};
    \end{axis}
    \end{tikzpicture}}
    \caption{Radar phase signal}
    \label{fig:full radar phase encoding muscle activity}
    \end{subfigure}
    
    \caption[During each experiment, two sets of time-synchronised radar and EMG data were recorded.]{During each experiment, a participant was asked to isometrically contract and relax their vastus lateralis muscle randomly. Two sets of time-synchronised data were then collected. These are radar and EMG data as presented here.}
    \label{fig:full emg, radar phase and camera phase encoding muscle activity}
\end{figure}
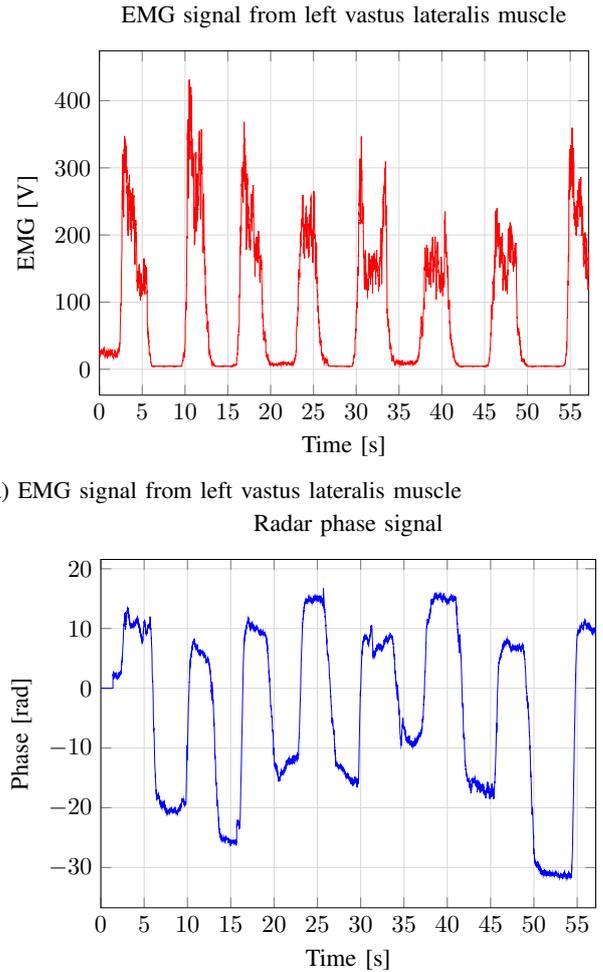

\begin{figure}[ht!]
    \begin{subfigure}[b]{\textwidth}
    \resizebox{0.55\textwidth}{!}{
    \begin{tikzpicture}
    \begin{axis}[title = { Normalised radar phase and EMG over a contraction-relaxation cycle},
    title style={text width=10cm, align=center},
    grid, 
    grid style={gray!30},
    xlabel = {Time [s]},
    ylabel = {},
    xmin=2,
    xmax=6,
    ymax=1.5, scale=4.5]
    \addplot[mark size=0.2, grid, color=red]table[y index=1]{participant1Exp1MAEMGRadarContractionCycleData.txt};
    \addplot[mark size=0.2, grid, color=blue]table[y index=2]{participant1Exp1MAEMGRadarContractionCycleData.txt};
    \legend {Normalised EMG, Normalised Radar Phase};
    \end{axis}
    \end{tikzpicture}
    }
    \caption{Normalised EMG and normalised radar phase}
    \label{fig:single contraction cycle of normalised emg and normalised radar phase against time}
    \end{subfigure}
    \begin{subfigure}[b]{\textwidth}
    \resizebox{0.55\textwidth}{!}{
    \begin{tikzpicture}
    \begin{axis}[title = {Normalised radar phase versus normalised EMG from the vastus lateralis},
    title style={text width=10cm, align=center},
    grid, 
    grid style={gray!30},
    xlabel = {Normalised EMG},
    ylabel = {Normalised Phase},
    ymin=0,
    ymax=1.4,
    xmin=0,
    xmax=1, scale=4.5]
    \addplot[only marks, mark size=0.2, grid, color=green]table[y index=1]{participant1Exp1MAEMGRadarBestFitExponentContractionCycleData.txt};
    \addplot[mark size=0.2, grid, color=black]table[y index=2]{participant1Exp1MAEMGRadarBestFitExponentContractionCycleData.txt};
    \legend {Data, Best Fit Line};
    \end{axis}
    \end{tikzpicture}
    } \captionsetup{justification=raggedright, singlelinecheck=false}
    \caption{Normalised radar phase versus normalised EMG and best 
    \newline fit line}
    \label{fig:normalised emg and normalised radar phase scatter plot over the contraction part of the contraction-relaxation cycle}
    \end{subfigure}
    \caption[The exponential relationship we observe between normalised EMG and the corresponding normalised radar phase from a single contraction-relaxation cycle is a typical characteristic of muscles.]{Illustrated in~(a)~is the normalised EMG and corresponding normalised radar phase from a single contraction-relaxation cycle in Fig.~\ref{fig:full emg, radar phase and camera phase encoding muscle activity}. The relationship observed between the two normalised signals is characteristic of muscles and, represented differently, it results in the hysteresis curve in Fig.~\ref{fig:on phase and off phase hysteresis curve}. A scatter plot of the normalised radar phase against normalised EMG during the contraction phase of~(a)~is presented in~(b). Also included in~(b)~is the exponential line that best fits the data. The exponential relationship observed here is typical. }
    \label{fig:normalised emg and radar scatter plot and time domain plot}
\end{figure}
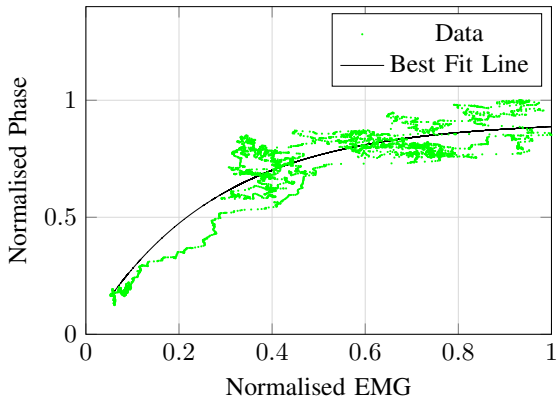

Another relationship that was observed between normalised EMG and normalised radar phase was that during contraction, the deformation was an exponential function of the EMG. The signals are normalised by dividing each sample by the largest sample in the signal. This exponential relationship was also observed in~\cite{Shi2008-ro} and~\cite{Hodges2003-ab} where deformation was measured through SMG. Fig.~\ref{fig:normalised emg and normalised radar phase scatter plot over the contraction part of the contraction-relaxation cycle} presents a scatter plot of normalised radar phase against normalised EMG for the first contraction in Fig.~\ref{fig:full emg, radar phase and camera phase encoding muscle activity}. The exponential relationship between the two variables can be seen in this plot.\\

Also depicted in Fig.~\ref{fig:normalised emg and normalised radar phase scatter plot over the contraction part of the contraction-relaxation cycle} is an exponential curve that best fits the data. Following~\cite{Shi2008-ro} and~\cite{Hodges2003-ab}, the best fit line was modelled as

\begin{equation}
\label{deformation_emg_eqn}
Y = A(1-e^{-BX}),
\end{equation}

where \(Y\) is the normalised muscle deformation, \(X\) is the normalised EMG, \(A\) is the asymptotic value of Y and \(B\) is the exponent coefficient determining the curvature.\\

To solve for the model coefficients \(A\) and \(B\), the problem was posed as a non-linear least-squares curve fitting problem and solved using the Levenberg–Marquardt algorithm~\cite{mathworksLeastSquaresModel}. For the data in Fig.~\ref{fig:normalised emg and normalised radar phase scatter plot over the contraction part of the contraction-relaxation cycle}, the values of \(A\) and \(B\) were found to be \(0.91\) and \(3.68\), respectively. The equation of the depicted best fit line is thus \(Y = 0.91(1-e^{-3.68X})\).\\

The model coefficients were calculated for all 12 experiments. Fig.~\ref{fig:model coefficients A and B across participants and experiments} below presents the 12 values of each coefficient across participants and experiments. The average values of A were found to be \(1\), \(0.91\) and \(1.06\) for participants~1, 2 and 3 respectively. For B, the average values were found to be \(3.45\), \(12.83\) and \(3.60\) for participants~1, 2 and 3 respectively. Based on~\cite{Shi2008-ro} and~\cite{Hodges2003-ab}, these are typical values. In~\cite{Shi2008-ro}, the average value of B for the biceps brachii muscle over seven participants is~\(6.19\). In~\cite{Hodges2003-ab} it is~\(7.69\) for the biceps brachii muscle over five participants whereas we have~\(6.63\) as the average value of B over all three participants. \\

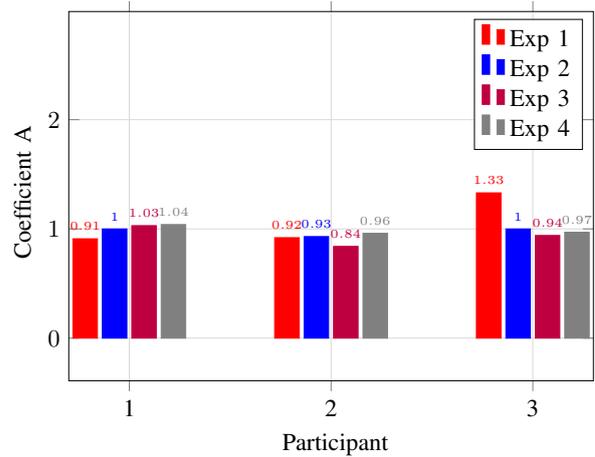
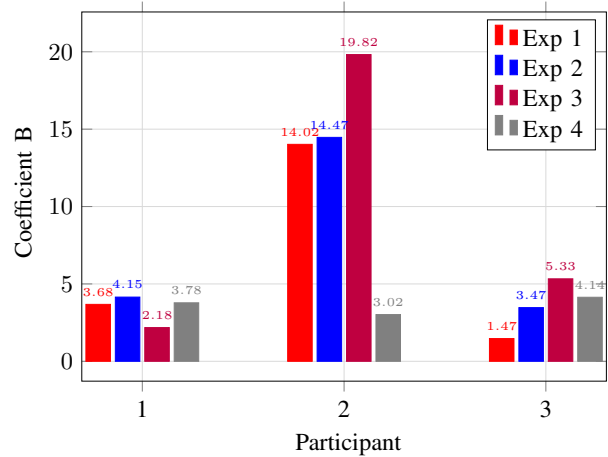
\begin{figure}[ht!]
    \begin{subfigure}[b]{\textwidth}
    \resizebox{0.52\textwidth}{!}{
    \begin{tikzpicture}
    \begin{axis}[title=Value of coefficient A across participants and experiments,
    title style={text width=10cm, align=center},
    ybar, 
    grid, 
    grid style={gray!30},
    enlargelimits=0.15,  
    ylabel={\ Coefficient A},
    xlabel={\ Participant}, 
    symbolic x coords={1, 2, 3},  
    xtick=data,  
    nodes near coords,  
    nodes near coords align={vertical}, 
    yticklabel style={/pgf/number format/fixed},
    ymax=2.6,
    ymin=0.0,
    scale=5.3]  
    \addplot [color=red, font=\tiny, fill=red] coordinates {(1,0.91) (2,0.92) (3,1.33)}; 
    \addplot [color=blue, font=\tiny, fill=blue] coordinates {(1,1) (2,0.93) (3,1)};
    \addplot [color=purple, font=\tiny, fill=purple] coordinates {(1,1.03) (2,0.84) (3,0.94)};
    \addplot [color=gray, font=\tiny, fill=gray] coordinates {(1,1.04) (2,0.96) (3,0.97)};
    \legend {Exp 1, Exp 2, Exp 3, Exp 4};
    \end{axis} 
    \end{tikzpicture}}
    \caption{Value of coefficient A across participants and experiments}
    \label{fig:model coefficients A across participants and experiments}
    \end{subfigure}
    \begin{subfigure}[b]{\textwidth}
    \resizebox{0.52\textwidth}{!}{
    \begin{tikzpicture}
    \begin{axis}[title=Value of coefficient B across participants and experiments,
    title style={text width=10cm, align=center},
    ybar, 
    grid, 
    grid style={gray!30},
    enlargelimits=0.15,  
    ylabel={\ Coefficient B}, 
    xlabel={\ Participant},  
    symbolic x coords={1, 2, 3},  
    xtick=data,  
    nodes near coords,  
    nodes near coords align={vertical}, 
    yticklabel style={/pgf/number format/fixed},
    scale=5.3]  
    \addplot [color=red, font=\tiny, fill=red] coordinates {(1,3.68) (2,14.02) (3,1.47)}; 
    \addplot [color=blue, font=\tiny, fill=blue] coordinates {(1,4.15) (2,14.47) (3,3.47)};
    \addplot [color=purple, font=\tiny, fill=purple] coordinates {(1,2.18) (2,19.82) (3,5.33)};
    \addplot [color=gray, font=\tiny, fill=gray] coordinates {(1,3.78) (2,3.02) (3,4.14)};
    \legend {Exp 1, Exp 2, Exp 3, Exp 4};
    \end{axis} 
    \end{tikzpicture}}
    \caption{Value of coefficient B across participants and experiments}
    \label{fig:model coefficients B across participants and experiments}
    \end{subfigure}
    
    \caption[MATLAB's lsqnonlin() method from the Optimisation Toolbox was used to solve a curve-fitting problem to find values for coefficients A and B in the exponential equation.]{The relationship of normalised radar phase to normalised EMG during the contraction of a muscle has been observed to be exponential, as depicted in Fig.~\ref{fig:normalised emg and normalised radar phase scatter plot over the contraction part of the contraction-relaxation cycle}. \eqref{deformation_emg_eqn} is often used to model the relationship and we do the same here. In~(a)~is the value of coefficient A and in~(b)~is the value of coefficient B for the data collected from all the experiments viewed per participant. MATLAB's lsqnonlin() method from the Optimisation Toolbox was used to fit \eqref{deformation_emg_eqn} to the data. }
    \label{fig:model coefficients A and B across participants and experiments}
\end{figure}

In Fig.~\ref{fig:model coefficients A and B across participants and experiments} we present the values of A and B for each participant separately, because the value of B may be fundamentally specific to each participant. Indeed, in~\cite{Shi2008-ro} it was established through Analysis of Variance (ANOVA) that there was a significant difference in B across the seven participants. Differences in B across participants means that the percentage change in muscle deformation per percentage change in the driving EMG varies across participants. The relatively high average B value of \(12.83\) for participant~2 (see Fig.~\ref{fig:model coefficients B across participants and experiments}) suggests that participant 2 has the highest percentage change in muscle deformation per percentage change in the driving EMG. Of course, this then means that the muscle deformation for this participant saturates at lower percentage values of the EMG. \\  

\begin{figure}[ht!]
    \begin{subfigure}[b]{\textwidth}
    \resizebox{0.52\textwidth}{!}{
    \begin{tikzpicture}
    \begin{axis}[title = {Observed radar phase and modelled radar phase},
    title style={text width=10cm, align=center},
    grid, 
    grid style={gray!30},
    align=center,
    xlabel = {Time [s]},
    ylabel = {},
    xmin=2,
    xmax=5.4993,
    ymax=1.4,
    scale=5] 
    \addplot[mark size=0.2, grid, color=red]table[y index=1]{participant1Exp1MAObservedModelledRadarDataDownSampled.txt};
    \addplot[mark size=0.2, grid, color=blue]table[y index=2]{participant1Exp1MAObservedModelledRadarDataDownSampled.txt};
    \legend {Observed Radar Phase, Modelled Radar Phase};
    \end{axis}
    \end{tikzpicture}
    }
    \caption{Observed and modelled radar phase}
    \label{fig:observed and modelled radar phase}
    \end{subfigure}
    \begin{subfigure}[b]{\textwidth}
    \resizebox{0.52\textwidth}{!}{
    \begin{tikzpicture}[scale=1,line width=1pt]
    \begin{axis}  
    [title=Coefficient of determination between EMG and radar phase,
    title style={text width=10cm, align=center},
    ybar,
    grid, 
    grid style={gray!30},
    enlargelimits=0.15,
    align=center,
    ylabel={Mean Coefficient of Determination},
    ylabel style={align=center, text width=5.3cm},
    xlabel={\ Participant},  
    symbolic x coords={1, 2, 3}, 
    xtick=data,
    ymin=0.6,
    ymax=0.8,
    nodes near coords, 
    nodes near coords align={vertical}, 
    yticklabel style={/pgf/number format/fixed},
    scale=5]  
    \addplot coordinates {(1,0.77) (2,0.63) (3,0.75)}; 
    \end{axis} 
    \end{tikzpicture}
    }
    \caption{Coefficient of determination between EMG and radar phase}
    \label{fig:coefficient of determination across participants}
    \end{subfigure}
    
    \caption[Using \eqref{deformation_emg_eqn} and the model parameters in Fig.~\ref{fig:model coefficients A and B across participants and experiments}, we computed the normalised radar phase predicted by the model given some normalised EMG from our data.]{Using~\eqref{deformation_emg_eqn} and the model parameters in Fig.~\ref{fig:model coefficients A and B across participants and experiments}, we computed the normalised radar phase predicted by the model given some normalised EMG from our data. In~(a)~is the resulting normalised radar phase and the normalised radar phase actually observed from the experiment in Fig.~\ref{fig:full emg, radar phase and camera phase encoding muscle activity}. With these computed radar phases, we computed the goodness of fit of our exponential models for all experiments. In~(b)~is the goodness of fit across participants. }
    \label{fig:observed and modelled radar phase and coefficient of determination across participants}
\end{figure}
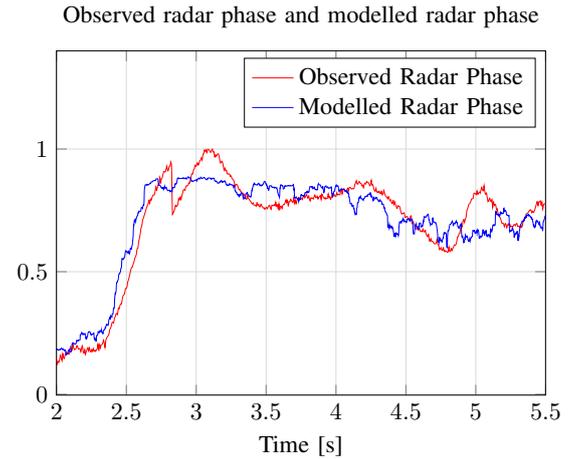
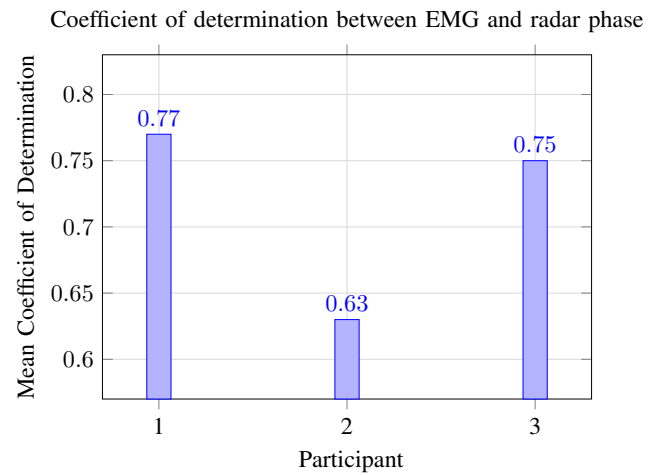

Once average values of A and B have been computed for a participant, \eqref{deformation_emg_eqn} can then be used as a predictive model. Given a normalised value of EMG from the participant, the normalised deformation that should result from this excitation can be computed. This was done using participant~1's parameters and a segment of normalised EMG data from the experiment depicted in Fig.~\ref{fig:full emg, radar phase and camera phase encoding muscle activity} as input to the model. Fig.~\ref{fig:observed and modelled radar phase} shows the modelled or predicted normalised radar phase and the
normalised radar phase actually observed from the experiment. \\

We have established that the relationship between radar phase and EMG is exponential (see Fig.~\ref{fig:normalised emg and normalised radar phase scatter plot over the contraction part of the contraction-relaxation cycle}). Now that has been established, we measure the goodness of fit of our exponential model to the observed data. We do this by computing the coefficient of determination \(R^2\). The coefficient of determination is a measure of the proportion of the variation in the dependent variable (the muscle deformation) that is predictable from the independent variable (the EMG) using our exponential model. The value of \(R^2\) is calculated as follows:

\begin{equation}
\label{coefficient_of_determination_eqn}
R^2 = 1 - \frac{\sum_{i=1}^{n} (y_i - f_i)^2}{\sum_{i=1}^{n}(y_i - \Bar{y})^2},
\end{equation}

where \(n\) is the total number of samples in the data used, \(y_i\) is the \(i\)th sample of the observed normalised radar phase, \(\Bar{y}\) is the mean value of the observed normalised radar phase and \(f_i\) is the \(i\)th sample of the normalised radar phase predicted by the model.\\

Eq.~\eqref{coefficient_of_determination_eqn} outputs a value between 0 and 1. The goodness of fit for the illustrative experiment in Fig.~\ref{fig:normalised emg and radar scatter plot and time domain plot} is \(0.88\). This means that \(88\%\) of the muscle deformation is predicted or accounted for by the EMG. Fig.~\ref{fig:coefficient of determination across participants} presents the average value of \(R^2\) for each of the three participants, averaged over the four experiments. We found the average value of \(R^2\) across all participants to be \(72\%\) whereas in~\cite{Shi2008-ro} they found the average value of \(R^2\) to be \(88\%\).\\

Another interesting phenomenon or relationship between EMG and muscle deformation can be observed in Fig.~\ref{fig:single contraction cycle of normalised emg and normalised radar phase against time}. The plots shown in Fig.~\ref{fig:single contraction cycle of normalised emg and normalised radar phase against time} can be considered in two stages, the ON or contraction stage from roughly \(2\)~s to \(3\)~s and the OFF or relaxation stage from roughly \(5.5\)~s to \(6\)~s. Notice that a given percentage of muscle deformation (normalised radar phase) corresponds to a lower percentage EMG during the OFF stage than during the ON stage. This phenomenon is easily observable when the normalised EMG is plotted against the normalised radar phase for only the ON and OFF stages. Fig.~\ref{fig:on phase and off phase hysteresis curve} below depicts such a plot, drawn from the data in Fig.~\ref{fig:single contraction cycle of normalised emg and normalised radar phase against time}. The resultant curve is called a hysteresis curve and was also observed in~\cite{Shi2008-ro} and~\cite{Orizio2003-qp}. In~\cite{Shi2008-ro}, the hysteresis was between muscle deformation and EMG and is exactly as we observe here. In~\cite{Orizio2003-qp} however, it was observed between muscle deformation and force and it was stated that such hysteresis curves are well known in biomechanics. The EMG amplitude increases during the ON stage and decreases during the OFF stage, and therefore the hysteresis in Fig.~\ref{fig:on phase and off phase hysteresis curve} (and~\cite{Shi2008-ro}) has a clockwise direction while that in~\cite{Orizio2003-qp} has an anti-clockwise direction.\\

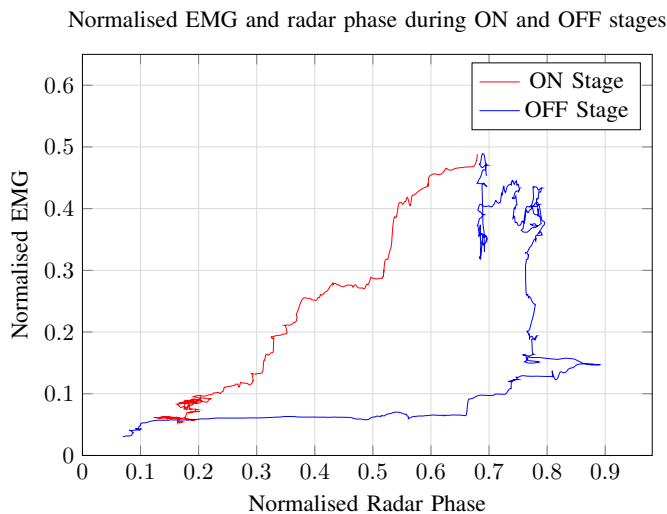
\begin{figure}[ht!]
    \resizebox{0.52\textwidth}{!}
    {
    \begin{tikzpicture}
    \begin{axis}[title = {Normalised EMG and radar phase during ON and OFF stages},
    title style={text width=10cm, align=center},
    grid, 
    grid style={gray!30},
    xlabel = {Normalised Radar Phase},
    ylabel = {Normalised EMG},
    ytick = {0, 0.1, 0.2, 0.3, 0.4, 0.5, 0.6},
    xtick = {0, 0.1, 0.2, 0.3, 0.4, 0.5, 0.6, 0.7, 0.8, 0.9, 1.0},
    ymax=0.65, 
    ymin=0.0,
    xmin=0.0,
    scale=6]
    \addplot[mark size=0.2, grid, color=red]table[y index=1]{participant1Exp1MAONPhaseRadarEMGData.txt};
    \addplot[mark size=0.2, grid, color=blue]table[y index=1]{participant1Exp1MAOFFPhaseRadarEMGData.txt};
    \legend{ON Stage, OFF Stage};
    \end{axis}
    \end{tikzpicture}
    }
    \caption[Presented here is the normalised EMG data from the illustrative experiment and illustrative contraction-relaxation cycle plotted against the corresponding normalised radar phase.]{Presented here is the normalised EMG data from the illustrative experiment and illustrative contraction-relaxation cycle plotted against the corresponding normalised radar phase. Only the ON and OFF stages of the contraction-relaxation cycle are presented. The part of the cycle when the contraction is held is not depicted. The hysteresis curve observed here is characteristic and well known in biomechanics~\cite{Shi2008-ro}~\cite{Orizio2003-qp}.}
    \label{fig:on phase and off phase hysteresis curve}
\end{figure}

The final observation common across our study and the studies in~\cite{Shi2008-ro} and~\cite{Orizio2003-qp} was that the rate of muscle deformation was typically higher during the ON stage compared to the OFF stage. As in~\cite{Shi2008-ro} however, this effect was not always observed. It was almost exclusively observed only in participant~1. In~\cite{Shi2008-ro}, this limitation was attributed to the low frame rate (\(8\)~Hz) of the ultrasound imaging system and the inability of the participants to reduce torque smoothly. In our study, the muscle deformation was sampled at the slow time frequency of \(178.5\)~Hz, which is more than 20 times that in~\cite{Shi2008-ro}. Because of the relatively higher sampling frequency of our system, this limitation is probably due to participant 2 and 3's inability to relax the vastus lateralis smoothly. \\

Additionally, it was noticed that the data from participant~1 contained significantly less random body motion as compared to participants~2 and 3. RBM is any motion of the body that is not the motion of interest. This RBM may also be the reason why participants~2 and 3 had lower average values of the coefficient of determination than participant~1 (see Fig.~\ref{fig:coefficient of determination across participants}). This is plausible since that would mean some of the variation in the muscle deformation that is unaccounted for by EMG is accounted for by the RBM. In future work, we plan on using the stereo camera data to compensate for RBM. \\


Finally, another limitation of our approach was that the recovered muscle deformation was prone to drift. Meaning that after contraction, the radar phase did not return back to \(0\)~rad even though the muscle was relaxed (see Fig.~\ref{fig:full emg, radar phase and camera phase encoding muscle activity}). This happens because after every contraction, the muscle does not return to exactly the same position relative to the radar-camera rig as it was before the contraction. Though an effort was made by each participant to avoid translation of the leg, the leg was not strapped in place. The SMG approach is also prone to this drift problem as demonstrated in~\cite{Hallock2021-au},~\cite{Kamatham2022-fc} and~\cite{9224391}. Methods for correcting this drift for ultrasound recovered muscle deformation exist~\cite{Kamatham2022-fc}. It was hypothesised in~\cite{Hallock2021-au} that drift was the cause of diminished performance and it was demonstrated in~\cite{9224391} that correcting this drift can improve some performance measures. Similarly, it was observed in our study that contraction stages severely affected by drift have very poor coefficients of determination.\\

Based on the foregoing results and discussions, we argue that the recovered radar phase signal is the same muscle deformation signal recovered through SMG. Several studies~\cite{Shi2008-ro}~\cite{9224391}~\cite{Kamatham2022-fc}~\cite{Hallock2021-au} ~\cite{Orizio2003-qp} have demonstrated that this signal correlates with muscle force. More importantly, experiments performed in~\cite{Hallock2021-au} have demonstrated that participants can be instructed to use this signal to track a trajectory. This deformation-based trajectory tracking can be performed with accuracy similar to that achieved with EMG-based trajectory tracking and participants prefer using this deformation signal. The muscle deformation signal is thus a new avenue for device control~\cite{Hallock2021-au}~\cite{Zheng2006}, e.g. the control of prosthetics.\\ 

However, unlike SMG, our approach is not only non-invasive but also non-contact. SMG is technically challenging to implement without contact because of the difference between the acoustic impedance of air and that of human tissue. At the interface between media with different acoustic impedances, some of the acoustic energy is transmitted into the other medium while some is reflected~\cite{REGTIEN2018267}. To reduce the amount of reflected energy at the air-skin interface, the acoustic impedance of the skin must match that of the media the ultrasound transducer is in (i.e. the two media must be acoustically coupled). To achieve this, an ultrasound gel is often applied on the skin and the transducer placed in it~\cite{Shi2008-ro}, making it impossible for the transducer to be ‘non-contact’ in any real sense.\\

\section{Conclusion}
\label{sec:conclusion}

The current state of the art for monitoring muscle activity is EMG. EMG is contact based, uncomfortable, time consuming, and requires a good understanding of anatomy. Our radar based non-contact muscle activity estimation approach addresses all these challenges and costs \(\$640\) compared to more than \(\$25~000\) for an EMG system. In this paper, we presented the theory of a radar based small motion monitoring approach. We demonstrated how this theory can be used to estimate muscle activity without contact. To the best of the authors' knowledge, our approach is the first to measure the characteristic dimensional changes of muscles in vivo and without contact. \\

For future work, we plan on extending our approach to enable RBM cancellation and the tracking and monitoring of
small motions for mobile participants. If the varying range
of the target is known, say from a calibrated stereo-camera pair, then the radar phase signal can be computed with the beat frequency, \(f_b\), updated using the range information. Applications of this include studying the gaits of people in the lab as well as animals in the wild. 

\bibliographystyle{IEEEtran}
\typeout{}
\bibliography{ajsen}

\begin{thebibliography}{10}
\providecommand{\url}[1]{#1}
\csname url@samestyle\endcsname
\providecommand{\newblock}{\relax}
\providecommand{\bibinfo}[2]{#2}
\providecommand{\BIBentrySTDinterwordspacing}{\spaceskip=0pt\relax}
\providecommand{\BIBentryALTinterwordstretchfactor}{4}
\providecommand{\BIBentryALTinterwordspacing}{\spaceskip=\fontdimen2\font plus
\BIBentryALTinterwordstretchfactor\fontdimen3\font minus
  \fontdimen4\font\relax}
\providecommand{\BIBforeignlanguage}[2]{{%
\expandafter\ifx\csname l@#1\endcsname\relax
\typeout{** WARNING: IEEEtran.bst: No hyphenation pattern has been}%
\typeout{** loaded for the language `#1'. Using the pattern for}%
\typeout{** the default language instead.}%
\else
\language=\csname l@#1\endcsname
\fi
#2}}
\providecommand{\BIBdecl}{\relax}
\BIBdecl

\bibitem{Chowdhury2013-yx}
R.~H. Chowdhury, M.~B.~I. Reaz, M.~A. B.~M. Ali, A.~A.~A. Bakar, K.~Chellappan,
  and T.~G. Chang, ``\BIBforeignlanguage{en}{Surface electromyography signal
  processing and classification techniques},''
  \emph{\BIBforeignlanguage{en}{Sensors (Basel)}}, vol.~13, no.~9, pp.
  12\,431--12\,466, Sep. 2013.

\bibitem{Woodward2019-cz}
R.~B. Woodward, M.~J. Stokes, S.~J. Shefelbine, and R.~Vaidyanathan,
  ``\BIBforeignlanguage{en}{Segmenting mechanomyography measures of muscle
  activity phases using inertial data},'' \emph{\BIBforeignlanguage{en}{Sci.
  Rep.}}, vol.~9, no.~1, p. 5569, Apr. 2019.

\bibitem{8462887}
L.~A. Hallock, A.~Kato, and R.~Bajcsy, ``Empirical quantification and modeling
  of muscle deformation: Toward ultrasound-driven assistive device control,''
  in \emph{2018 IEEE International Conference on Robotics and Automation
  (ICRA)}, 2018, pp. 1825--1832.

\bibitem{Hallock2021-au}
L.~A. Hallock, B.~Sud, C.~Mitchell, E.~Hu, F.~Ahamed, A.~Velu, A.~Schwartz, and
  R.~Bajcsy, ``\BIBforeignlanguage{en}{Toward real-time muscle force inference
  and device control via optical-flow-tracked muscle deformation},''
  \emph{\BIBforeignlanguage{en}{IEEE Trans. Neural Syst. Rehabil. Eng.}},
  vol.~29, pp. 2625--2634, Dec. 2021.

\bibitem{9707638}
K.~Li, M.~Tucker, R.~Gehlhar, Y.~Yue, and A.~D. Ames, ``Natural multicontact
  walking for robotic assistive devices via musculoskeletal models and hybrid
  zero dynamics,'' \emph{IEEE Robotics and Automation Letters}, vol.~7, no.~2,
  pp. 4283--4290, 2022.

\bibitem{9025819}
Y.~Masuda and M.~Ishikawa, ``Autonomous intermuscular coordination and leg
  trajectorygeneration of neurophysiology-based quasi-quadruped robot,'' in
  \emph{2020 IEEE/SICE International Symposium on System Integration (SII)},
  2020, pp. 1123--1128.

\bibitem{Merletti2016-ez}
R.~Merletti and D.~Farina, Eds., \emph{\BIBforeignlanguage{en}{Surface
  Electromyography}}, ser. IEEE Press Series on Biomedical Engineering.\hskip
  1em plus 0.5em minus 0.4em\relax Nashville, TN: John Wiley \& Sons, Apr.
  2016.

\bibitem{Orizio1993-vg}
C.~Orizio, ``\BIBforeignlanguage{en}{Muscle sound: bases for the introduction
  of a mechanomyographic signal in muscle studies},''
  \emph{\BIBforeignlanguage{en}{Crit. Rev. Biomed. Eng.}}, vol.~21, no.~3, pp.
  201--243, 1993.

\bibitem{Barry1985}
\BIBentryALTinterwordspacing
D.~T. Barry, S.~R. Geiringer, and R.~D. Ball, ``Acoustic myography: A
  noninvasive monitor of motor unit fatigue,'' \emph{Muscle \& Nerve}, vol.~8,
  no.~3, pp. 189--194, Mar. 1985. [Online]. Available:
  \url{https://doi.org/10.1002/mus.880080303}
\BIBentrySTDinterwordspacing

\bibitem{DBLP:conf/memea/CasacciaSCTR15}
\BIBentryALTinterwordspacing
S.~Casaccia, L.~Scalise, L.~Casacanditella, E.~P. Tomasini, and J.~W.
  Rohrbaugh, ``Non-contact assessment of muscle contraction: {L}aser {D}oppler
  {M}yography,'' in \emph{2015 {IEEE} International Symposium on Medical
  Measurements and Applications, MeMeA 2015, Torino, Italy, May 7-9,
  2015}.\hskip 1em plus 0.5em minus 0.4em\relax {IEEE}, 2015, pp. 610--615.
  [Online]. Available: \url{https://doi.org/10.1109/MeMeA.2015.7145276}
\BIBentrySTDinterwordspacing

\bibitem{Rohrbaugh2013-mz}
J.~W. Rohrbaugh, E.~J. Sirevaag, and E.~J. Richter,
  ``\BIBforeignlanguage{en}{Laser {D}oppler vibrometry measurement of the
  mechanical myogram},'' \emph{\BIBforeignlanguage{en}{Rev. Sci. Instrum.}},
  vol.~84, no.~12, p. 121706, Dec. 2013.

\bibitem{Shi2008-ro}
J.~Shi, Y.-P. Zheng, Q.-H. Huang, and X.~Chen,
  ``\BIBforeignlanguage{en}{Continuous monitoring of sonomyography,
  electromyography and torque generated by normal upper arm muscles during
  isometric contraction: sonomyography assessment for arm muscles},''
  \emph{\BIBforeignlanguage{en}{IEEE Trans. Biomed. Eng.}}, vol.~55, no.~3, pp.
  1191--1198, Mar. 2008.

\bibitem{Kamatham2022-fc}
A.~T. Kamatham, M.~Alzamani, A.~Dockum, S.~Sikdar, and B.~Mukherjee, ``A
  simple, drift compensated method for estimation of isometric force using
  sonomyography,'' in \emph{Sensing Technology}, ser. Lecture notes in
  electrical engineering.\hskip 1em plus 0.5em minus 0.4em\relax Cham: Springer
  International Publishing, 2022, pp. 355--366.

\bibitem{s22072789}
\BIBentryALTinterwordspacing
L.~Brausch, H.~Hewener, and P.~Lukowicz, ``Classifying muscle states with
  one-dimensional radio-frequency signals from single element ultrasound
  transducers,'' \emph{Sensors}, vol.~22, no.~7, 2022. [Online]. Available:
  \url{https://www.mdpi.com/1424-8220/22/7/2789}
\BIBentrySTDinterwordspacing

\bibitem{seniamWelcomeSENIAM}
``{W}elcome to {S}{E}{N}{I}{A}{M} --- seniam.org,''
  \url{http://www.seniam.org/}, [Accessed 14-Jun-2023].

\bibitem{Manca2020-ez}
A.~Manca, A.~Cereatti, L.~Bar-On, A.~Botter, U.~Della~Croce, M.~Knaflitz, N.~A.
  Maffiuletti, D.~Mazzoli, A.~Merlo, S.~Roatta, A.~Turolla, and F.~Deriu,
  ``\BIBforeignlanguage{en}{A survey on the use and barriers of surface
  electromyography in neurorehabilitation},''
  \emph{\BIBforeignlanguage{en}{Front. Neurol.}}, vol.~11, p. 573616, Oct.
  2020.

\bibitem{517ff2062ffd4c67869d303afd50d7f7}
C.~Orizio, M.~Gobbo, A.~Veicsteinas, R.~Baratta, B.~Zhou, and M.~Solomonow,
  ``\BIBforeignlanguage{English}{Transients of the force and surface
  mechanomyogram during cat gastrocnemius tetanic stimulation},''
  \emph{\BIBforeignlanguage{English}{European Journal of Applied Physiology}},
  vol.~88, no.~6, pp. 601--606, Feb. 2003.

\bibitem{REGTIEN2018267}
\BIBentryALTinterwordspacing
P.~Regtien and E.~Dertien, ``9 - acoustic sensors,'' in \emph{Sensors for
  Mechatronics (Second Edition)}, 2nd~ed., P.~Regtien and E.~Dertien,
  Eds.\hskip 1em plus 0.5em minus 0.4em\relax Elsevier, 2018, pp. 267--303.
  [Online]. Available:
  \url{https://www.sciencedirect.com/science/article/pii/B9780128138106000094}
\BIBentrySTDinterwordspacing

\bibitem{Haim_Azhari_2010-qo}
H.~Azhari, ``Special imaging techniques,'' in \emph{Basics of Biomedical
  Ultrasound for Engineers}.\hskip 1em plus 0.5em minus 0.4em\relax Hoboken,
  NJ, USA: John Wiley \& Sons, Inc., Apr. 2010, p. 313.

\bibitem{9224391}
L.~A. Hallock, A.~Velu, A.~Schwartz, and R.~Bajcsy, ``Muscle deformation
  correlates with output force during isometric contraction,'' in \emph{2020
  8th IEEE RAS/EMBS International Conference for Biomedical Robotics and
  Biomechatronics (BioRob)}, 2020, pp. 1188--1195.

\bibitem{McMahon1984-rk}
T.~A. McMahon, \emph{\BIBforeignlanguage{en}{Muscles, reflexes, and
  locomotion}}.\hskip 1em plus 0.5em minus 0.4em\relax Princeton, NJ: Princeton
  University Press, Apr. 1984.

\bibitem{Chen2019-wh}
V.~C. Chen, \emph{The micro-Doppler effect in radar the micro-Doppler effect in
  radar}, 2nd~ed.\hskip 1em plus 0.5em minus 0.4em\relax Norwood, MA: Artech
  House, Feb. 2019.

\bibitem{heartRateSensingICRA2020}
P.~Zhao, C.~X. Lu, B.~Wang, C.~Chen, L.~Xie, M.~Wang, N.~Trigoni, and
  A.~Markham, ``Heart rate sensing with a robot mounted mm{W}ave radar,'' in
  \emph{International Conference on Robotics and Automation (ICRA)}, 2020.

\bibitem{Adib2015-ld}
F.~Adib, H.~Mao, Z.~Kabelac, D.~Katabi, and R.~C. Miller, ``Smart homes that
  monitor breathing and heart rate,'' in \emph{Proceedings of the 33rd Annual
  {ACM} Conference on Human Factors in Computing Systems}.\hskip 1em plus 0.5em
  minus 0.4em\relax New York, NY, USA: ACM, Apr. 2015.

\bibitem{8123923}
Y.~Xiong, Z.~Peng, G.~Xing, W.~Zhang, and G.~Meng, ``Accurate and robust
  displacement measurement for {FMCW} radar vibration monitoring,'' \emph{IEEE
  Sensors Journal}, vol.~18, no.~3, pp. 1131--1139, 2018.

\bibitem{6662489}
C.~Gu, G.~Wang, Y.~Li, T.~Inoue, and C.~Li, ``A hybrid radar-camera sensing
  system with phase compensation for random body movement cancellation in
  doppler vital sign detection,'' \emph{IEEE Transactions on Microwave Theory
  and Techniques}, vol.~61, no.~12, pp. 4678--4688, 2013.

\bibitem{imbalanceCorr1}
B.-K. Park, S.~Yamada, and V.~Lubecke, ``Measurement method for imbalance
  factors in direct-conversion quadrature radar systems,'' \emph{IEEE Microwave
  and Wireless Components Letters}, vol.~17, no.~5, pp. 403--405, 2007.

\bibitem{imbalanceCorr2}
A.~Singh, X.~Gao, E.~Yavari, M.~Zakrzewski, X.~H. Cao, V.~M. Lubecke, and
  O.~Boric-Lubecke, ``Data-based quadrature imbalance compensation for a {CW}
  {D}oppler radar system,'' \emph{IEEE Transactions on Microwave Theory and
  Techniques}, vol.~61, no.~4, pp. 1718--1724, 2013.

\bibitem{9716146}
Z.~Xu, C.~Shi, T.~Zhang, S.~Li, Y.~Yuan, C.-T.~M. Wu, Y.~Chen, and
  A.~Petropulu, ``Simultaneous monitoring of multiple people’s vital sign
  leveraging a single phased-{MIMO} radar,'' \emph{IEEE Journal of
  Electromagnetics, RF and Microwaves in Medicine and Biology}, vol.~6, no.~3,
  pp. 311--320, 2022.

\bibitem{Alizadeh2019RemoteMO}
M.~Alizadeh, G.~Shaker, J.~C.~M. de~Almeida, P.~P. Morita, and
  S.~Safavi-Naeini, ``Remote monitoring of human vital signs using mm{W}ave
  {FMCW} radar,'' \emph{IEEE Access}, vol.~7, pp. 54\,958--54\,968, 2019.

\bibitem{tiIWR6843ISKEvaluation}
``{I}{W}{R}6843{I}{S}{K} {E}valuation board | {T}{I}.com --- ti.com,''
  \url{https://www.ti.com/tool/IWR6843ISK}, [Accessed 14-Jun-2023].

\bibitem{tiDCA1000EVMEvaluation}
``{D}{C}{A}1000{E}{V}{M} {E}valuation board | {T}{I}.com --- ti.com,''
  \url{https://www.ti.com/tool/DCA1000EVM}, [Accessed 14-Jun-2023].

\bibitem{tiMMWAVEICBOOSTEvaluation}
``{M}{M}{W}{A}{V}{E}{I}{C}{B}{O}{O}{S}{T} {E}valuation board | {T}{I}.com ---
  ti.com,'' \url{https://www.ti.com/tool/MMWAVEICBOOST}, [Accessed
  14-Jun-2023].

\bibitem{Hodges2003-ab}
P.~W. Hodges, L.~H.~M. Pengel, R.~D. Herbert, and S.~C. Gandevia,
  ``\BIBforeignlanguage{en}{Measurement of muscle contraction with ultrasound
  imaging},'' \emph{\BIBforeignlanguage{en}{Muscle Nerve}}, vol.~27, no.~6, pp.
  682--692, Jun. 2003.

\bibitem{mathworksLeastSquaresModel}
``{L}east-{S}quares ({M}odel {F}itting) {A}lgorithms - {M}{A}{T}{L}{A}{B} \&;
  {S}imulink - {M}ath{W}orks {U}nited {K}ingdom --- uk.mathworks.com,''
  \url{https://uk.mathworks.com/help/optim/ug/least-squares-model-fitting-algorithms.html},
  [Accessed 14-Jun-2023].

\bibitem{Orizio2003-qp}
C.~Orizio, M.~Gobbo, A.~Veicsteinas, R.~V. Baratta, B.~H. Zhou, and
  M.~Solomonow, ``\BIBforeignlanguage{en}{Transients of the force and surface
  mechanomyogram during cat gastrocnemius tetanic stimulation},''
  \emph{\BIBforeignlanguage{en}{Eur. J. Appl. Physiol.}}, vol.~88, no.~6, pp.
  601--606, Feb. 2003.

\bibitem{Zheng2006}
\BIBentryALTinterwordspacing
Y.~Zheng, M.~Chan, J.~Shi, X.~Chen, and Q.~Huang, ``Sonomyography: Monitoring
  morphological changes of forearm muscles in actions with the feasibility for
  the control of powered prosthesis,'' \emph{Medical Engineering \& Physics},
  vol.~28, no.~5, pp. 405--415, Jun. 2006. [Online]. Available:
  \url{https://doi.org/10.1016/j.medengphy.2005.07.012}
\BIBentrySTDinterwordspacing

\end{thebibliography}

\begin{IEEEbiography}[{\includegraphics[width=1in,height=1.25in,clip,keepaspectratio]{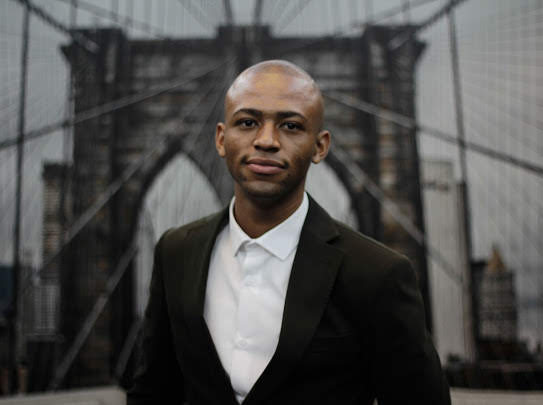}}]{Kukhokuhle Tsengwa} received the B.Sc. degree in mechatronics engineering from the University of Cape Town (UCT) in 2020. He is set to graduate from the M.Sc. degree in electrical engineering from UCT in 2023. He is currently pursuing the PhD degree at the Oxford Robotics Institute at the University of Oxford working on control barrier functions for obstacle avoidance for walking robots and supported by the Rhodes Trust. His research interests include control, dynamical systems and signal processing.
\end{IEEEbiography}

\begin{IEEEbiography}[{\includegraphics[width=1in,height=1.25in,clip,keepaspectratio]{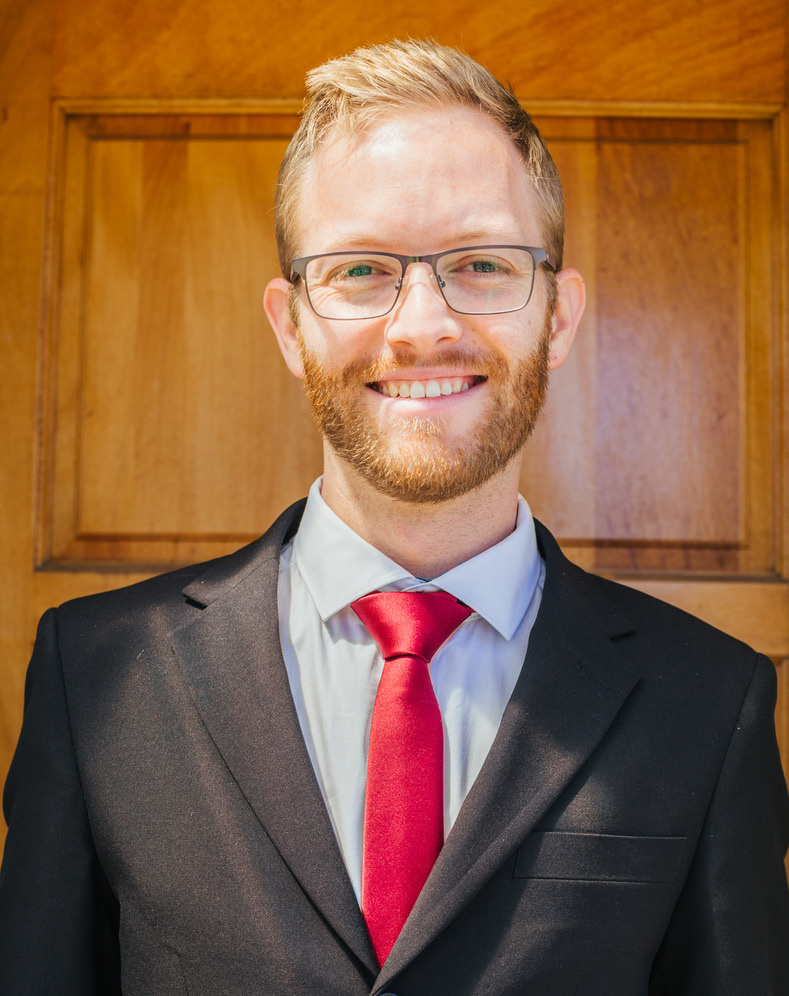}}]{Stephen Paine} holds an M.Sc. (Eng) in radar and electronic defense specializing in RF and antenna design, and a Ph.D. (Eng) specializing in passive radar and electronic warfare from the University of Cape Town. He is currently a Lecturer with the Department of Electrical Engineering at the University of Cape Town, where he is also the Head of the Radar and Remote Sensing Group (RRSG). He has previously worked in industry in South Africa and Switzerland, with research interests in the field of radar, remote sensor fusion, signal processing, and automotive systems. He is a Junior Member of the IEEE.
\end{IEEEbiography}

\begin{IEEEbiography}[{\includegraphics[width=1in,height=1.25in,clip,keepaspectratio]{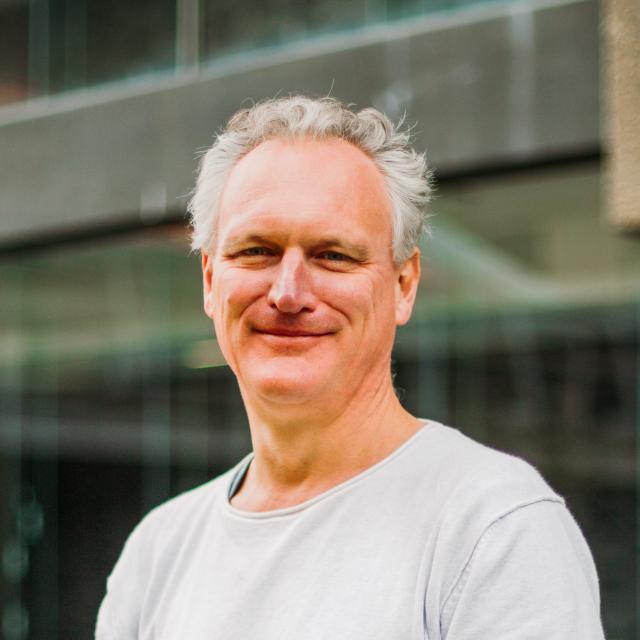}}]{Fred Nicolls} is an Associate Professor in the Department of Electrical Engineering at the University of Cape Town, Cape Town, South Africa.
\end{IEEEbiography}

\begin{IEEEbiography}[{\includegraphics[width=1in,height=1.25in,clip,keepaspectratio]{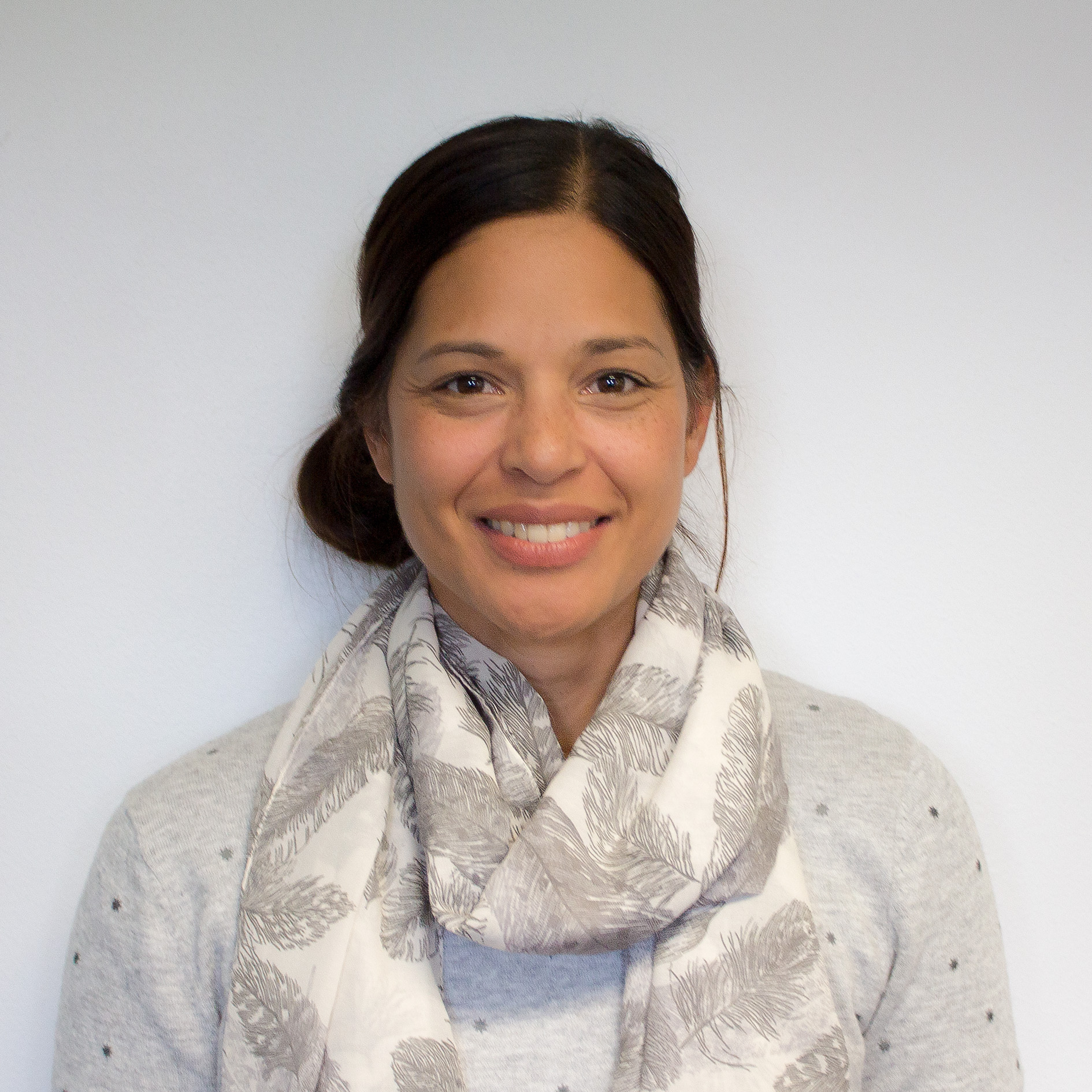}}]{Yumna Albertus} received a Ph.D Exercise Science in 2008 from the University of Cape Town. She is currently an Associate Professor with the Department of Human Biology, University of Cape Town. She is a research member of the Health, Physical Activity, Lifestyle and Sport Research Centre (HPALS) and UCT Neuroscience Institute. Her research interests include neuromuscular physiology (involving both bipolar and high density EMG) in performance and clinical conditions, specifically spinal cord injury and Cerebral Palsy. She was awarded UCT’s Research Leadership Award, one of 10 female researchers recognised for their impactful research. She is the current Chair of the South African Society of Biomechanics and the African representative on the International Society of Biomechanics Executive Board.
\end{IEEEbiography}

\begin{IEEEbiography}[{\includegraphics[width=1in,height=1.25in,clip,keepaspectratio]{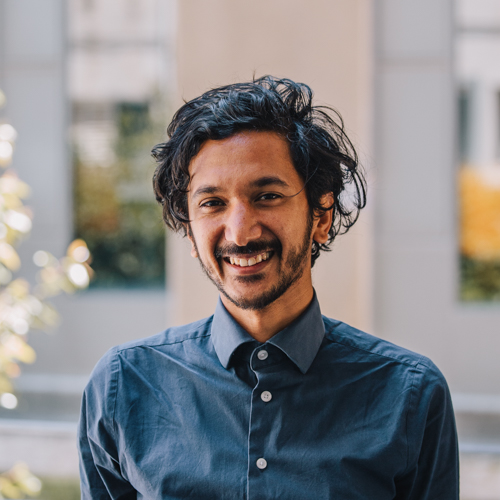}}]{Amir Patel} (Senior Member, IEEE) received the B.S. degree in mechatronics, the M.S. degree in control engineering, and the Ph.D. degree in mechatronics from the University of Cape Town, Cape Town, South Africa, in 2009, 2011, and 2014, respectively. He is currently an Associate Professor with the Department of Electrical Engineering, University of Cape Town, and the Director of the African Robotics Unit. He was previously a Visiting Research Scholar with Carnegie Mellon University, Pittsburgh, PA, USA, and Johns Hopkins University, Baltimore, MD, USA, and a Senior Software Developer with Tellumat, Cape Town, South Africa. His research interests include maneuverability in legged animals, bioinspired robotics, optimal control, and novel sensing systems. Dr. Patel was awarded the Claude Leon Merit Award in 2016 and then the Oppenheimer Memorial Trust Fellowship in 2018. He was a runner-up for the IEEE-RAS Technical Committee on Model-Based Optimization for Robotics Best Paper Award in 2019.
\end{IEEEbiography}

\end{document}